%% file: main.tex
\documentclass[10pt,conference]{IEEEtran}
\usepackage{graphicx}
\usepackage{multicol}
\usepackage{multirow}
\usepackage{listings}
\usepackage{url}
\usepackage{xcolor}
\usepackage{amsmath}
\usepackage{caption}
\usepackage{subcaption}
\usepackage{pifont}
\usepackage{float}

\begin{document}

\title{Performance Analysis and Optimization Opportunities for NVIDIA Automotive GPUs}

\author{
	\IEEEauthorblockN{
		Hamid Tabani\IEEEauthorrefmark{1},
		Fabio Mazzocchetti\IEEEauthorrefmark{1}\IEEEauthorrefmark{2},
		Pedro Benedicte\IEEEauthorrefmark{1}\IEEEauthorrefmark{2},
		Jaume Abella\IEEEauthorrefmark{1} and
		Francisco J. Cazorla\IEEEauthorrefmark{1}
	}
	\IEEEauthorblockA{
		\IEEEauthorrefmark{1} Barcelona Supercomputing Center
		\hspace{1em}
		\IEEEauthorrefmark{2} Universitat Polit\`{e}cnica de Catalunya
	}
}

	\maketitle

	\begin{abstract}
	Advanced Driver Assistance Systems (ADAS) and Autonomous Driving (AD) bring unprecedented performance requirements for automotive systems. Graphic Processing Unit (GPU) based platforms have been deployed with the aim of meeting these requirements, being NVIDIA Jetson TX2 and its high-performance successor, NVIDIA AGX Xavier, relevant representatives. However, to what extent high-performance GPU configurations are appropriate for ADAS and AD workloads remains as an open question.

	This paper analyzes this concern and provides valuable insights on this question by modeling two recent automotive NVIDIA GPU-based platforms, namely TX2 and AGX Xavier. In particular, our work assesses their microarchitectural parameters against relevant benchmarks, identifying GPU setups delivering increased performance within a similar cost envelope, or decreasing hardware costs while preserving original performance levels. Overall, our analysis identifies opportunities for the optimization of automotive GPUs to further increase system efficiency.
	\end{abstract}
	
	\linespread{0.99}
	\selectfont

	\input{1.0.Introduction.tex}

	\input{2.0.Background.tex}
	\input{3.0.Modelling.tex}

	\input{4.0.Design.tex}
	\input{5.0.Improvements.tex}

	\input{6.0.Related.tex}
	\input{7.0.Conclusions.tex}

	\section*{Acknowledgments}
	
	This work has been partially supported by the Spanish Ministry of Economy and Competitiveness (MINECO) under
	grant TIN2015-65316-P, the European Research Council (ERC) under the European Union's Horizon 2020 research and innovation programme (grant agreement No. 772773) and the HiPEAC Network of Excellence.
	
	\bibliographystyle{plain}
	\bibliography{biblio}

\end{document}

%% file: 1.0.Introduction.tex
\section{Introduction}
\label{sec:introduction}

Critical real-time embedded systems (CRTES), such as those managing safety-related functionalities in avionics, space, automotive and railway, have built during decades on simple and low-performance microcontrollers. The increasing software complexity, inherent to the increase in the number and sophistication of delivered functionalities in those systems, has lead towards a slow adoption of multicore microcontrollers. For instance, the Infineon AURIX processor family~\cite{infineon_aurix} in the automotive domain, or Gaisler's LEON4 family~\cite{website:ngmp} in the space domain deliver few cores (i.e. in the range 3 to 6) with the aim of providing a moderate performance scale up. {\color{white}\footnote{https://doi.org/10.1016/j.jpdc.2021.02.008}}

Such designs have already found difficulties to match the increasing complexity of software in those systems, which has increased at a rate of 10x every 10 years and, for the automotive domain reached up to 100 million lines of code for some cars in 2009~\cite{Char09}.
Moreover, as pointed out by ARM prospects, the advent of driver assistance systems and autonomous driving in the automotive domain will lead to a performance demand increase of 100x in the timeframe 2016-2024~\cite{ARM100xB}, thus further exacerbating the performance needs for CRTES.

The answer to this performance demand has been the deployment of accelerators along with the microcontrollers, being Graphic Processing Units (GPUs) the main representative of those~\cite{NVIDIA-ADAS,Intel-ADAS,pujol2019generating} despite the existing challenges using GPUs in this domain~\cite{tabani2019assessing,alcaide2018safety}. In particular, several products such as Renesas R-Car H3~\cite{RenesasRCarH3}, NVIDIA Jetson TX2~\cite{Tegra1KAutonomous} and NVIDIA Jetson Xavier~\cite{Xavier,xavier1} have already reached the market building upon GPU technology inherited from the high-performance domain. Automotive GPUs have inherited designs devised for the high-performance domain with the aim of reducing costs in the design, verification and validation process for chip manufacturers.

Unfortunately, reusability of high-performance hardware does not consider GPUs efficiency in the automotive domain and, to the best of our knowledge, the design space for GPUs, where resources are sized with the aim of optimizing specific goals such as performance, has not been yet thoroughly performed for the automotive domain.

This paper aims at covering this gap by providing a GPU design exploration for the automotive domain by analyzing the influence that different microarchitectural hardware parameters, such as cache sizes, number of streaming multiprocessors (SMs), and the like have on performance for an automotive SoC representative, the NVIDIA Jetson TX2 platform~\cite{Tegra1KAutonomous} and AGX Xavier~\cite{xavier1}. In particular, the main contributions of this work are as follows:
\begin{enumerate}
    \item An adaptation of a cycle-level CPU-GPU simulator, gem5-gpu~\cite{power2015gem5}, to model both NVIDIA Jetson TX2 and AGX Xavier configurations as close as possible to the specs provided by NVIDIA. This is the basis of our exploratory study.
    \item A systematic performance analysis for the main hardware parameters, assessing to what extent they influence performance for a relevant set of benchmarks. We first analyze the effect of each parameter individually, and then the effect of two or more parameters coordinately.
    \item Finally, we propose two configurations which deliver: a) similar performance to the baseline design at a decreased hardware costs, and b) higher performance than the baseline design with comparable hardware costs.
\end{enumerate}
Overall, our analysis shows that opportunities for optimization exist and can be exploited in two different axes depending on the needs of end-users.

The rest of the paper is organized as follows. Section~\ref{sec:background} provides some background on automotive systems needing GPUs and on GPU architecture. Section~\ref{sec:modelling} presents our methodology to model the NVIDIA Jetson TX2 GPU, NVIDIA AGX Xavier and the benchmarks we use to evaluate our proposals. Section~\ref{sec:design} provides the design space exploration for the Jetson TX2 GPU and AGX Xavier. Section~\ref{sec:improved} identifies and evaluates our two improved setups. Section~\ref{sec:related} reviews some related work, and Finally, Section~\ref{sec:conclusions} summarizes the main conclusions of this paper.

%% file: 2.0.Background.tex
\section{Background}
\label{sec:background}

This section describes some background on the need for GPUs in automotive systems, as well as the basic organization of latest GPU architectures.

\subsection{GPU-based Automotive Systems}

The advent of Advanced Driver Assistance Systems and Autonomous Driving (AD) imposes a higher level of system autonomy to take decisions on behalf of the driver, or even to fully replace the driver, as expected for the systems with Autonomy Level 5 -- the highest autonomy level according to SAE International~\cite{J3016}. 
To make these systems real, a number of processes related to \textit{Perception} and \textit{Prediction} modules of autonomous driving systems need to be automated to process large amounts of data from sensors (camera, LiDAR, radar) in real-time to deliver system responses timely~\cite{apolloAuto,udacity,autoware}. Therefore, object detection, trajectory prediction, and collision avoidance algorithms, among others, use complex deep learning models, which are extremely compute-intensive and require very high-performance hardware to take driving decisions in very short timeframes.

{\color{black}
Current trends in the sophisticated AD systems show that the number of concurrent deep learning instances can easily reach dozens. The main reasons for such an increase are the following:
\begin{itemize}
	\item \textit{More input sensors}. Moving towards fully autonomous driving (Level 5~\cite{sae_levels}) naturally requires
	increasing the number of sensors to cover the car’s surrounding more accurately. Today, some
	of the AD systems, which are still far from a Level 5 system, use more than 8 cameras and
	radars (e.g., Telsa~\cite{tesla_autopilot} uses 8 cameras and 12 ultrasonic sensors, and NVIDIA autopilot~\cite{nvidiaReport}
	uses 8 high-resolution cameras, 8 radars and optionally up to 3 LiDARs). Therefore, more
	DNN-based workloads have to be processed, increasing the computation demand significantly.
	\item \textit{More sophisticated algorithms}. Perception submodules tend to use more sophisticated DNNs,
	with a larger number of layers and higher computational needs for further improvements in
	the accuracy of object and obstacle detection, especially in conditions with reduced visibility
	such as fog, dusk, night, rain, and snow. The Prediction module already uses 3 different neural
	networks, either to achieve higher accuracy or to cover more complex scenarios. Indeed, this
	type of module usually uses sophisticated neural network architectures~\cite{zyner2018recurrent,zyner2020naturalistic}.
	\item \textit{More functionalities}. Besides the main functions of an AD system, extra features are introduced to improve driving quality and safety: from gesture detection and speech-based command and control, up to driver-monitoring to predict take-over readiness~\cite{deo2020looking}.
\end{itemize}
}

\subsection{GPU Architecture}

\begin{figure}[t!]
\centering
\includegraphics[width=\columnwidth]{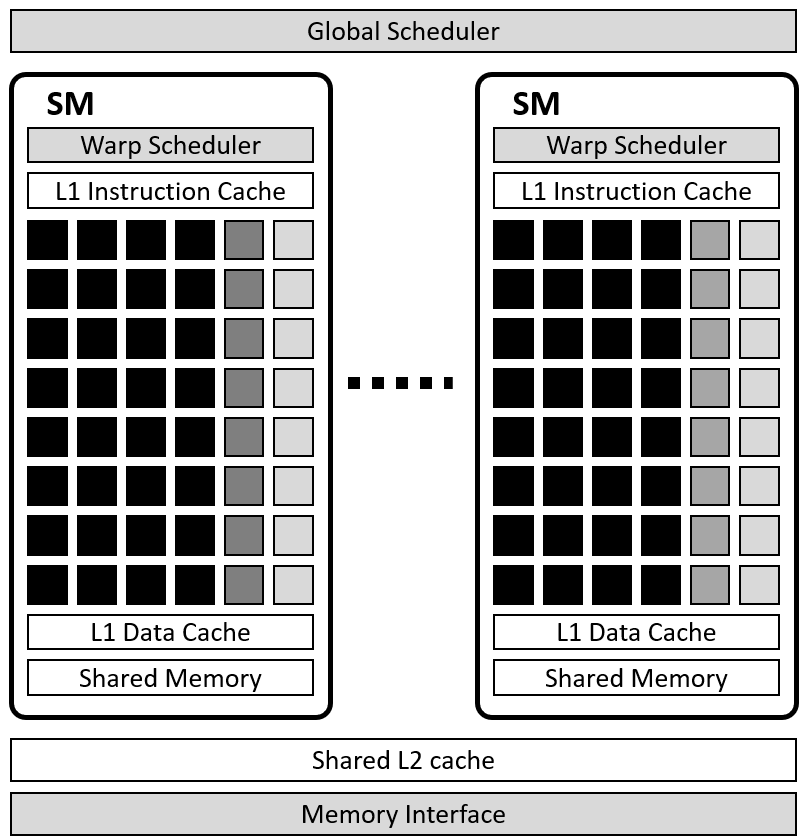}
\caption{Schematic of the architecture of a GPU.}
\label{fig:GPUarch}
\end{figure}

While different GPUs from different vendors may have significant differences, and differences may also be relevant across different GPU generations for the same vendor, some elements are mostly common to all GPUs. Therefore, we describe those, as they are the basis of the study provided later on.

Figure~\ref{fig:GPUarch} depicts the main elements of the architecture of the GPU. As the Figure shows, the GPU has a global scheduler that dispatches work to the different Streaming Multiprocessors (SMs) of the GPU. Each SM, to some extent, is a cluster of computing resources orchestrated coordinately, whereas different SMs may lack any coordination and work highly independently. Each SM has a set of storage resources that include a first level (L1) instruction cache and an L1 data cache, as well as some shared memory to manage shared data. Each SM also includes a \emph{warp} scheduler, where warp refers, in NVIDIA terminology, to the set of identical computations that are dispatched to the parallel computing elements atomically. Thus, it is the smallest scheduling unit.

{\color{black}
Computing elements include CUDA\footnote{Compute Unified Device Architecture (CUDA) is a parallel computing platform and application programming interface model created by NVIDIA. It allows software developers and software engineers to use a CUDA-enabled graphics processing unit for general purpose processing.} cores (indicated with black squares in the figure), able to process a warp entirely, as long as the particular operation requested is part of the CUDA cores, which typically include most integer and single-precision floating-point arithmetic operations. Along with the CUDA cores, some other units (indicated with different gray squares) perform other types of operations, typically with lower bandwidth than CUDA cores, such as load/store operations of data, and some area-costly operations whose hardware cannot be replicated as many times as CUDA cores (e.g., double precision or highly-complex floating point operations).

SMs also share one or more levels of cache (e.g., L2), thus competing for cache space across SMs, as well as the memory interface. Note, however, that different GPU architectures may be different. For instance, it is not uncommon having clusters of SMs, so that each cluster shares a cluster-local shared cache, and then all clusters share a global L2 cache. Still, the concept of highly parallel computing resources and hierarchical storage organization holds in the general case for GPUs.

NVIDIA TX2 GPU is based on Pascal Architecture~\cite{nvidia2016p100}. The GPU in AGX Xavier is designed based on the Volta~\cite{nvidia2017v100} architecture, the successor of Pascal architecture, with new design features and new computing elements such as Tensor cores~\cite{tensorcore} for accelerating deep learning-based kernels as Figure~\ref{fig:Voltaarch} shows.
Tensor cores accelerate large matrix operations, which are at the heart of many AI functions~\cite{hamidISORC2020}. While each regular core can perform up to one single-precision multiply-accumulate operation per 1 GPU clock, each Tensor core can perform one matrix multiply-accumulate operation per 1 GPU clock. The Tensor core can multiply two FP16 4x4 matrices and adds the multiplication product FP32 matrix to the accumulator, which is also a FP32 4x4 matrix. In each SM, threads can use either the regular cores or the Tensor cores. Hence, at most, 512 regular or 64 Tensor cores can be used in parallel.

\begin{figure}[t!]
	\centering
	\includegraphics[width=\columnwidth]{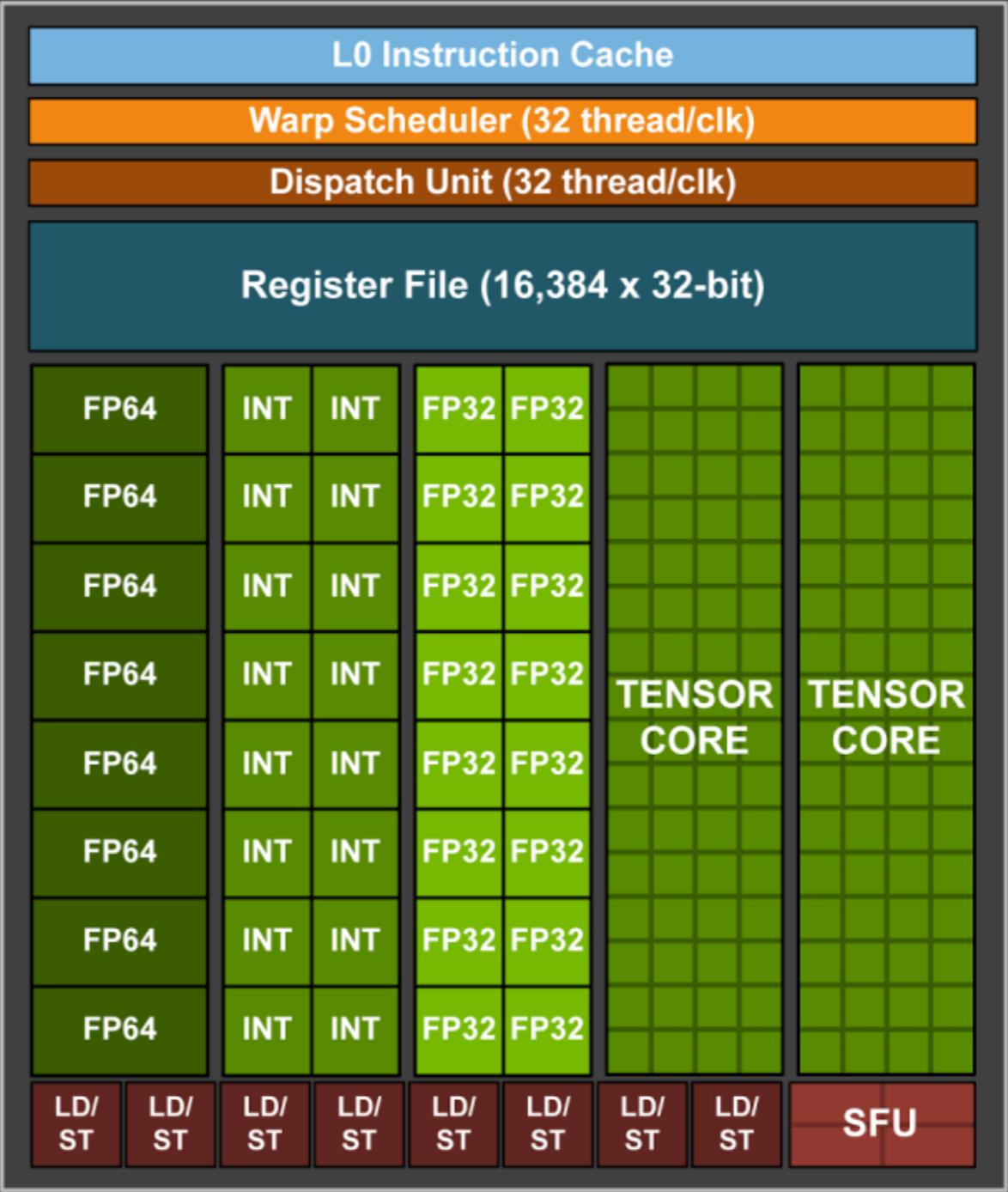}
	\caption{Schematic of the SM architecture of a Volta-based GPU~\cite{nvidia2017v100}.}
	\label{fig:Voltaarch}
\end{figure}
}

%% file: 3.0.Modelling.tex
\section{Methodology}
\label{sec:modelling}

\subsection{Simulation Infrastructure}

In this paper, we have used an in-house version of the gem5-gpu simulator~\cite{power2015gem5}.
Gem5-gpu is a cycle-accurate and heterogeneous CPU-GPU simulator incorporating gem5~\cite{binkert2011gem5} and gpgpu-sim~\cite{bakhoda2009analyzing} simulators. These are the most accurate and widely-used cycle-level simulators in the computer architecture community.
In our version of gem5-gpu, we applied major modifications to include latest versions of gem5 and gpgpu-sim. This was a fundamental requirement before being able to model latest GPU architectures such as Pascal and Volta since in the baseline simulator only Fermi, an earlier NVIDIA architecture, was supported.

\subsection{Modeling NVIDIA Tegra X2}

We have modeled an SoC, similar to NVIDIA TX2, in our simulator.
We have done a comprehensive study to extract the available and public architectural parameters of TX2 in order to tune the simulator to closely model the chosen platform.
We have extracted as much information as we could from public sources about our NVIDIA GPU and we included them in the configuration of the gpgpu-sim. 
However, we still do not have access to part of the detailed parameters since they are not provided by the manufacturer. Alternatively, we have tried to estimate the missing parameters according to the available information and also by fine-tuning those parameters with the help of synthetic experiments.
Table~\ref{tab:simulator-tx2} shows the detailed parameters that we have employed in our simulator to model the NVIDIA TX2.

To out knowledge, the latest gpgpu-sim version (released in late 2018) is the most accurate and open-source simulator to model a Pascal architecture, which is the architecture used in the TX2 GPU. In addition, we have designed several synthetic benchmarks to validate our configuration against the real GPU.

NVIDIA TX2 is based on the 16nm NVIDIA Tegra Parker system on chip (SoC). The TX2 has a Pascal GPU with 2 SMs, each of them with 4 SM Blocks (SMBs). Each SMB comprises of 32 cores and in total the GPU has 256 cores. The TX2 SoC also comprises of two different clusters of dual- and quad-core CPUs, whose L2 cache is shared inside each cluster. The first CPU cluster, \textit{Denver2}, has 2 cores each with its own private first level instruction and data caches (referred to as iL1 and dL1 respectively). The other CPU cluster has 4 ARM Cortex-A57 cores also with private iL1 and dL1 caches. Table~\ref{tab:tx2-parameters} presents the architectural parameters of both CPUs and the GPU in NVIDIA TX2. In this work, we focus on the development board of the TX2 processor (Jetson), which has one SoC. It is worth pointing out that commercial versions of the TX2 can have up to 2 SoCs like the one described and even one discrete GPU.

\definecolor{mygray2}{rgb}{0.85,0.85,0.85}
\begin{scriptsize}
\begin{table}[t!]
  \centering
  \small
  \begin{tabular}{|r|l|}
    \hline
     & \textbf{Configuration}\\
    \hline
	\hline
    GPU  & NVIDIA Pascal: 256 CUDA cores  \\
		 & 2 SMs with 4 SMBs each.  \\
		 & 32 CUDA cores per SMB \\ 
    \hline
	CPU	 	& 2-core Denver2 \\
	        & (128KB 4-way IL1, 64KB, 4way dL1)  \\
		   	& 4-core ARM A57 \\
		   	& (48KB 3-way IL1, 32KB 2-way dL1)  \\
		   	& 2MB 16-way L2 per cluster \\
	\hline
	DRAM & 8 GB, 256-bit LPDDR4x, 59.7 GB/s \\
	\hline
  \end{tabular}
  \caption{System Configurations of NVIDIA TX2 SoC.}
  \label{tab:tx2-parameters}
\end{table}
\end{scriptsize}

\begin{scriptsize}
\begin{table}[t!]
  \centering
  \small
  \begin{tabular}{|r|l|}
    \hline
     & \textbf{CPU Configuration}\\
    \hline
	\hline
    Core & ARMv8 ISA, 2.0 GHz, 128-entry ROB \\
		 & 40-entry Issue Queue, Full Out-of-Order  \\
		 & 3-Width Decoder, 3-Width Instruction Dispatch \\ 
		 & 48 KB, 3-Way TLB \\
		 & 48-Entry Fully-Associative L1 TLB \\
    \hline
	Caches & 32 KB L1-D Cache, 2-Way, 1 Cycle \\
		   & 48 KB L1-I Cache, 3-Way, 1 Cycle \\
		   & 2 MB L2 Cache, 16-Way, 12 Cycles \\
		   & 64 Bytes Cache Line Size \\
	\hline
	Prefetcher & Stride Prefetcher (Degree 1) \\
 			   & 2K Branch Target Buffer (BTB) \\
			   & 32-Instruction Fetch Queue \\
			   & 15 Cycles Misprediction Penalty \\
	\hline
	DRAM & DDR4 1866 MHz, 2 Ranks/Channel \\
		 & 8 Banks/Rank, 8 KB Row Size. \\
		 & \textit{t\textsubscript{CAS}} = \textit{t\textsubscript{RCD}} = \textit{t\textsubscript{RP}} = CL = 13.75ns \\
		 & \textit{t\textsubscript{REFI}} = 7.8 us \\
	\hline
	\hline
	& \textbf{GPU Configuration}\\
	\hline
	\hline
	SMs		& 1.1 GHz, 32 Warps, 65536 shader registers \\
			& 32 Thread blocks, 2048 threads per core \\
			& 4 scheduler per core \\
	\hline
	Memory	& 48KB 4-way, 512 KB 4-way L2 \\
			& 64KB shared memory, 8 GB total memory size \\
			& 16 sub-partition per memory channel \\ 
	\hline
  \end{tabular}
  \caption{System Configurations employed in the gem5-gpu simulator to model the NVIDIA TX2.}
  \label{tab:simulator-tx2}
\end{table}
\end{scriptsize}

{\color{black}
\subsection{Modelling NVIDIA AGX Xavier}

We have also modeled an SoC, similar to NVIDIA AGX Xavier, in our simulator.
Analogously to the case of the NVIDIA TX2, we have reviewed public information to extract the available and public architectural parameters of the AGX Xavier in order to tune the simulator to closely model the chosen platform through the configuration of the gpgpu-sim simulator.
However, similarly to the TX2 case, as explained before, we lack access to part of the detailed parameters since they are not publicly provided by the manufacturer. Alternatively, we have tried to estimate the missing parameters according to the available information and also by reverse-engineering those parameters with the help of synthetic experiments on the actual board.
Table~\ref{tab:simulator-xavier} shows the detailed parameters that we have employed in our simulator to model the AGX Xavier SoC.

To our knowledge, the latest gpgpu-sim version (released in late 2018), which we use in our work, is the most accurate open-source simulator to model a Volta architecture, which is the architecture used in the Xavier GPU.

NVIDIA Xavier SoC is based on the 12 nm FinFET system on chip (SoC). The Xavier GPU has a Volta GPU with 8 SMs, each of them with 4 SM Blocks (SMBs). Each SMB comprises of 16 cores and in total the GPU has 512 cores. The Xavier SoC comprises of four clusters of dual-core CPUs, 8 cores overall, whose L2 cache is shared inside each cluster and an L3 cache shared among the clusters. Each CPU cluster has 2 cores each with its own private first-level instruction and data caches (referred to as iL1 and dL1 respectively). Table~\ref{tab:xavier-parameters} presents the architectural parameters of both the CPU and the GPU in NVIDIA AGX Xavier. In this work, we focus on the development board of the Xavier (Jetson AGX Xavier), which has one single Xavier SoC. It is worth pointing out that commercial versions of the Xavier, similar to TX2, are designed with up to 2 Xavier SoCs in the same board.

\begin{scriptsize}
	\begin{table}[t!]
		\centering
		\small
		\begin{tabular}{|r|l|}
			\hline
			 & \textbf{Configuration}\\
			\hline
			\hline
			GPU  & NVIDIA Volta: 512 CUDA cores  \\
			& 8 SMs with 4 SMBs each.  \\
			& 16 CUDA cores per SMB \\ 
			\hline
			CPU	 	& 8-core ARMv8 \\ 
			& (128KB 4-way IL1, 64KB, 4way dL1)  \\
			& 2MB 16-way L2 per cluster \\
			& 4MB 16-way L3 shared between all clusters \\
			\hline
			DRAM & 32 GB, 256-bit LPDDR4x, 137 GB/s \\
			\hline
		\end{tabular}
		\caption{System Configurations of NVIDIA AGX Xavier SoC.}
		\label{tab:xavier-parameters}
	\end{table}
\end{scriptsize}

\begin{scriptsize}
	\begin{table}[t!]
		\centering
		\small
		\begin{tabular}{|r|l|}
			\hline
			 & \textbf{CPU Configuration}\\
			\hline
			\hline
			Core & ARMv8 ISA, 2.0 GHz, 128-entry ROB \\
			& 40-entry Issue Queue, Full Out-of-Order  \\
			& 3-Width Decoder, 3-Width Instruction Dispatch \\ 
			& 48 KB, 3-Way TLB \\
			& 48-Entry Fully-Associative L1 TLB \\
			\hline
			Caches & 64 KB L1-D Cache, 4-Way, 1 Cycle \\
			& 128 KB L1-I Cache, 4-Way, 1 Cycle \\
			& 2 MB L2 Cache, 16-Way, 12 Cycles \\
			& 4 MB L3 Cache 16-way, 16 Cycles \\
			& 64 Bytes Cache Line Size \\
			\hline
			Prefetcher & Stride Prefetcher (Degree 1) \\
			& 2K Branch Target Buffer (BTB) \\
			& 32-Instruction Fetch Queue \\
			& 15 Cycles Misprediction Penalty \\
			\hline
			DRAM & DDR4 1866 MHz, 2 Ranks/Channel \\
			& 8 Banks/Rank, 8 KB Row Size. \\
			& \textit{t\textsubscript{CAS}} = \textit{t\textsubscript{RCD}} = \textit{t\textsubscript{RP}} = CL = 13.75ns \\
			& \textit{t\textsubscript{REFI}} = 7.8 us \\
			\hline
			\hline
			 & \textbf{GPU Configuration}\\
			\hline
			\hline
			SMs		& 1.37 GHz, 32 Warps, 65536 shader registers \\
			& 32 Thread blocks, 2048 threads per core \\
			& 4 scheduler per core \\
			\hline
			Memory	& 64KB 4-way, 512 KB 4-way L2 \\
			&32 GB total memory size \\
			& 16 sub-partition per memory channel \\ 
			\hline
		\end{tabular}
		\caption{System Configurations employed in the gem5-gpu simulator to model the NVIDIA AGX Xavier.}
		\label{tab:simulator-xavier}
	\end{table}
\end{scriptsize}

}

\subsection{Benchmarks}

In this paper, we use Rodinia~\cite{rodinia} benchmark suite for our experiments. 
Rodinia benchmark suite is targeting heterogeneous computing and in order to study emerging platforms such as GPUs, Rodinia suite includes applications and kernels that target multi-core CPU and GPU platforms.

The EEMBC (Embedded Microprocessor Benchmarks Consortium) recently released ADASMark~\cite{adasmark}, an ADAS Benchmark suite that would be highly relevant for our study. However, in the moment of writing this paper we still have not been able to access this benchmark suite. Therefore, we used some of the most suitable benchmarks for GPU microarchitecture such as Rodinia. In fact, Rodinia includes some key kernels in autonomous driving systems that have similarities with ADASMark such as image processing and pattern recognition.

\begin{table*}[h]
	\centering
	\normalsize
	\begin{tabular}{|llll|}
		\hline
		\multicolumn{1}{|c}{\textbf{Name}} & \multicolumn{1}{c}{\textbf{Short name}} & \multicolumn{1}{c}{\textbf{Problem type}} & \multicolumn{1}{c|}{\textbf{Domain}} \\ \hline
		Back Propagation & backprop & Unstructured Grid & Pattern Recognition \\
		Breadth-First Search & bfs & Graph Traversal & Graph Algorithms \\
		3D Stencil & cell & Structured Grid & Cellular Automation \\
		Gaussian Elimination & gaussian & Dense Linear Algebra & Linear Algebra \\
		Hotspot3D & hotspot & Structured Grid & Physics Simulation \\
		Myocyte & myocyte & Structured Grid & Biological Simulation \\
		Needleman-Wunsch & needle & Dynamic Programming & Bioinformatics \\
		k-Nearest Neighbors & nn & Dense Linear Algebra & Data Mining \\
		Particle Filter & pf\_float &	Structured Grid & Medical Imaging \\
		Particle Filter & pf\_naive	& Structured Grid & Medical Imaging \\
		SRAD & srad & Structured Grid & Image Processing \\
		\hline
	\end{tabular}

	\caption{Rodinia benchmarks used in the experiments.}
	\label{fig:rodinia}
	\vspace{-0.2in}
\end{table*}

%% file: 4.0.Design.tex
\section{Design Space Exploration}
\label{sec:design}

The objective of the design space exploration is to know which parameters of the processor design could be increased/decreased and what would be their impact on performance. Since there are many parameters that can influence performance, making all the possible combinations is unfeasible. Thus, for each of the SoCs, we first change one parameter at a time, and then changing more than one parameter together (for instance, size and way of caches).

\begin{figure*}[h!]
	\centering

	\begin{subfigure}[b]{0.32\textwidth}
		\includegraphics[width=\textwidth]{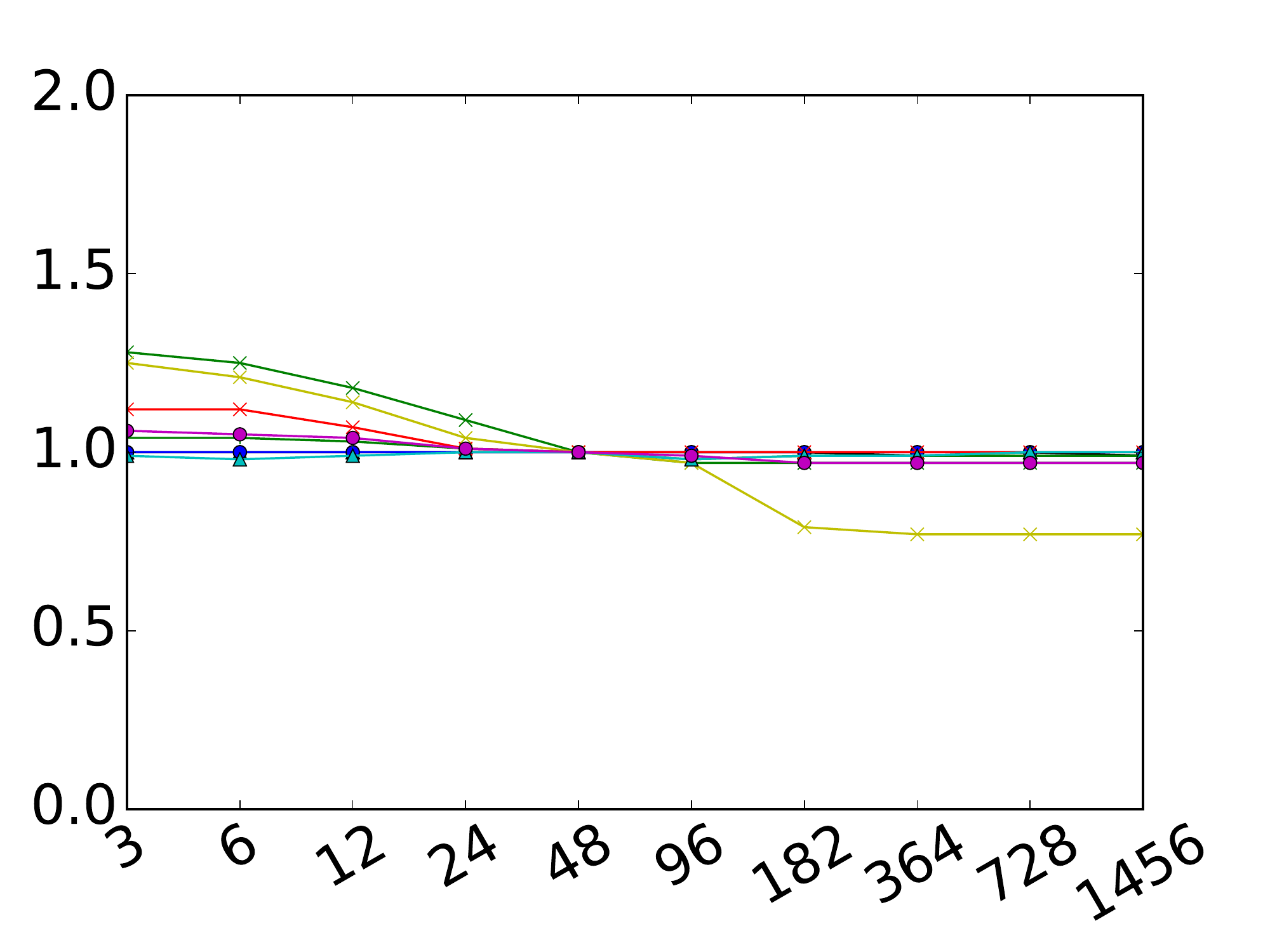}
		\caption{L1 size}
		\label{fig:dse_l1size}
	\end{subfigure}
	\begin{subfigure}[b]{0.32\textwidth}
		\includegraphics[width=\textwidth]{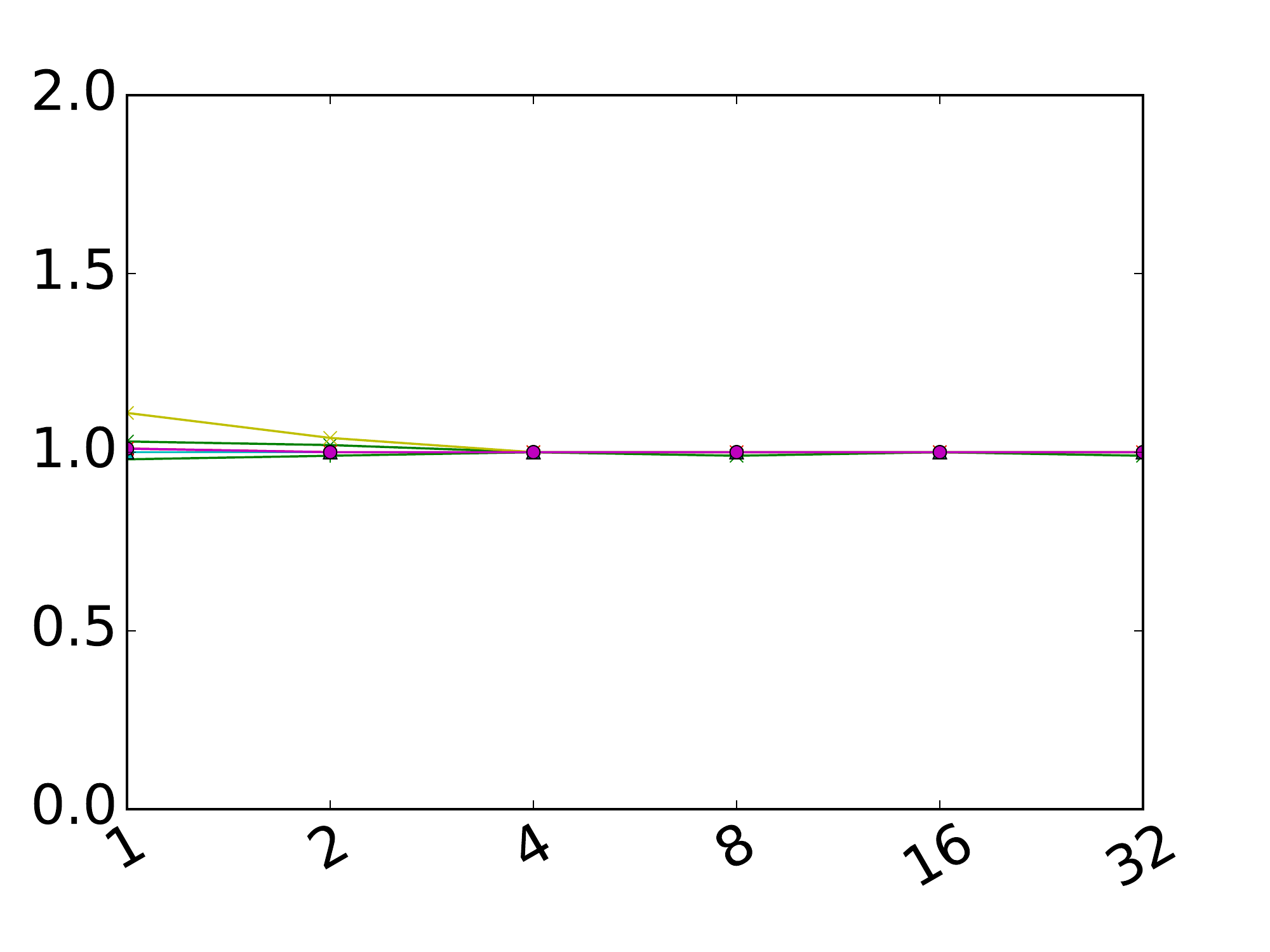}
		\caption{L1 associativity}
		\label{fig:dse_l1assoc}
	\end{subfigure}
	\begin{subfigure}[b]{0.32\textwidth}
		\includegraphics[width=\textwidth]{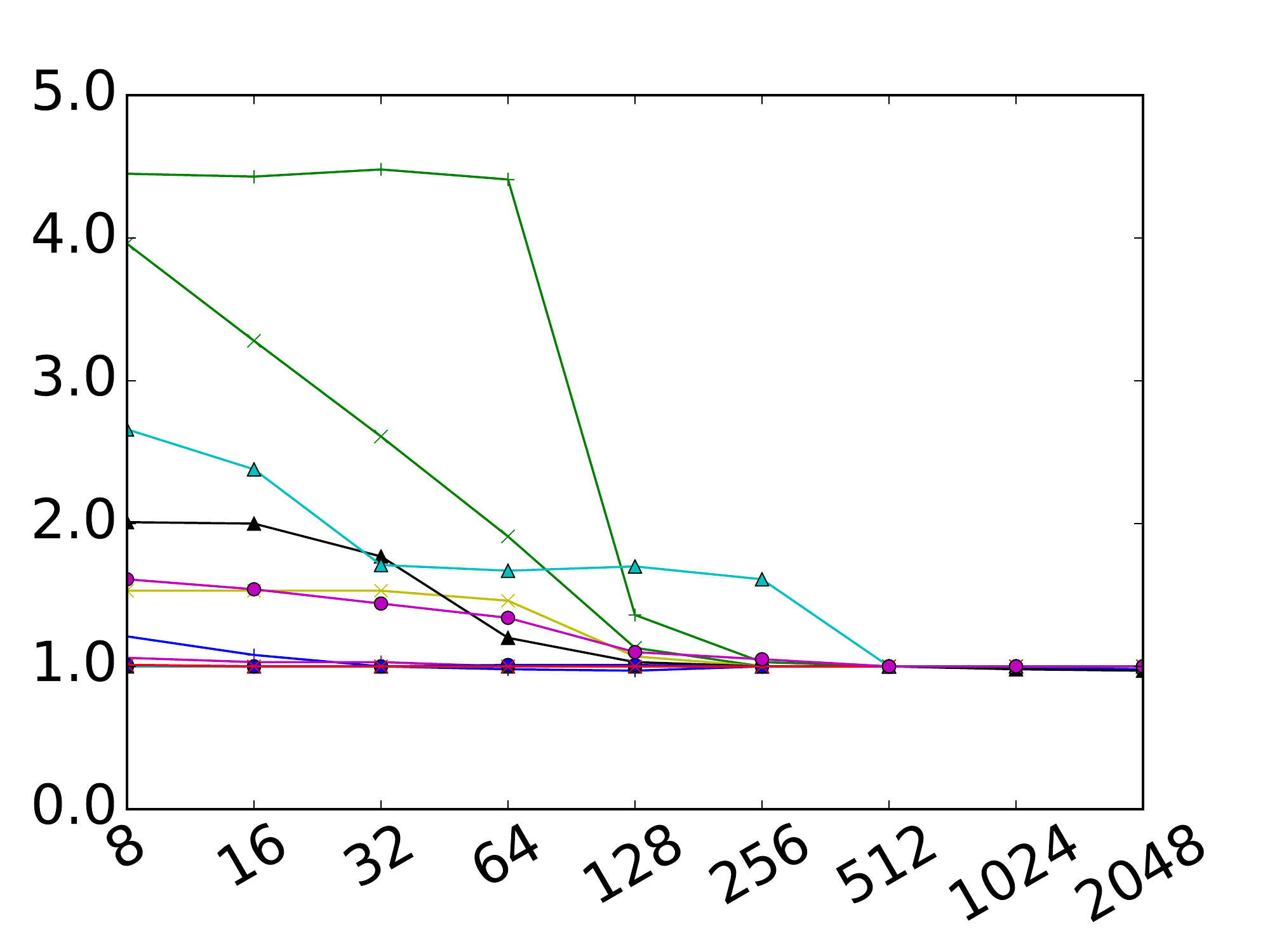}
		\caption{L2 size}
		\label{fig:dse_l2size}
	\end{subfigure}

	\begin{subfigure}[b]{0.32\textwidth}
		\includegraphics[width=\textwidth]{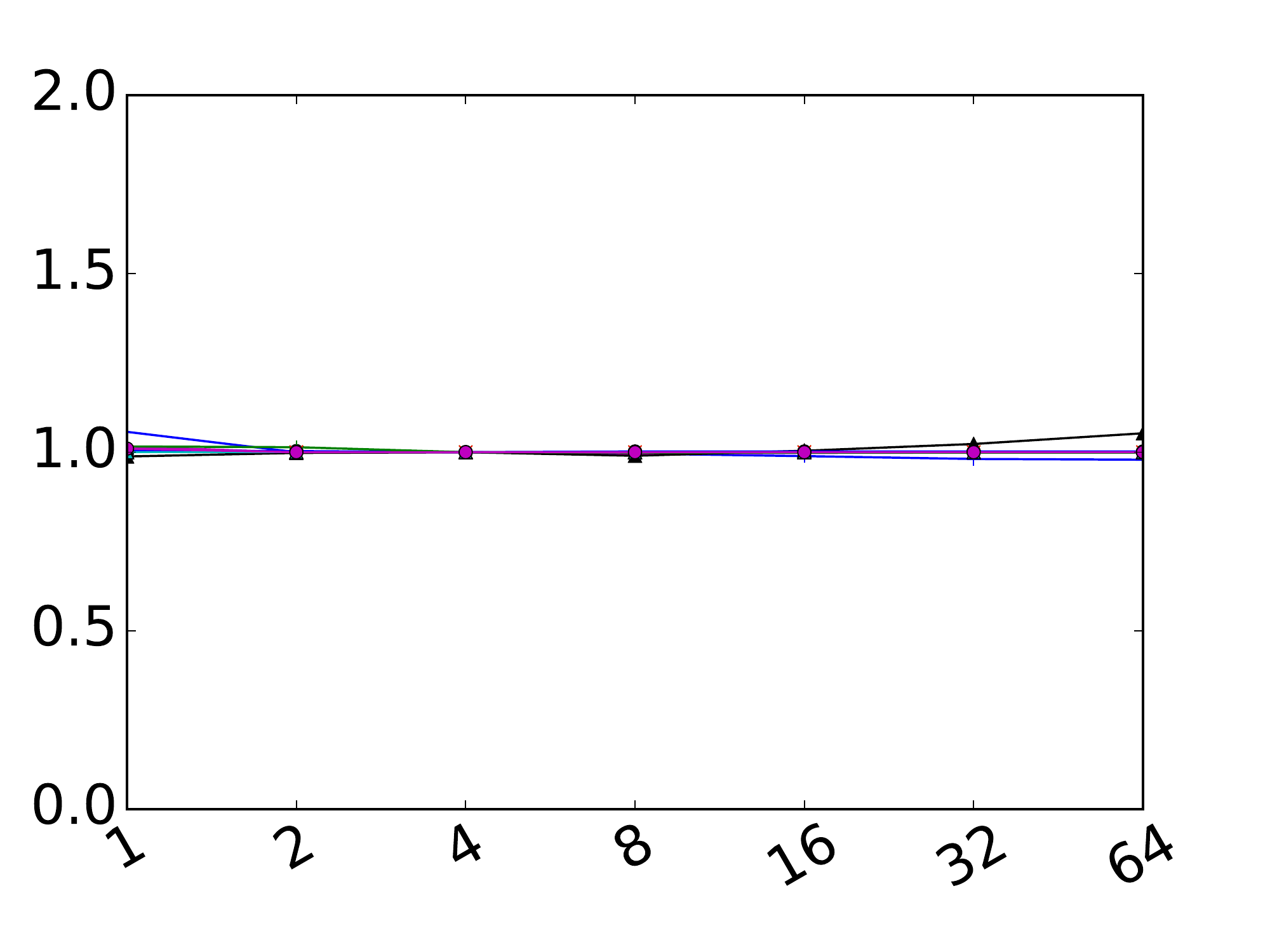}
		\caption{L2 associativity}
		\label{fig:dse_l2assoc}
	\end{subfigure}
	\begin{subfigure}[b]{0.32\textwidth}
		\includegraphics[width=\textwidth]{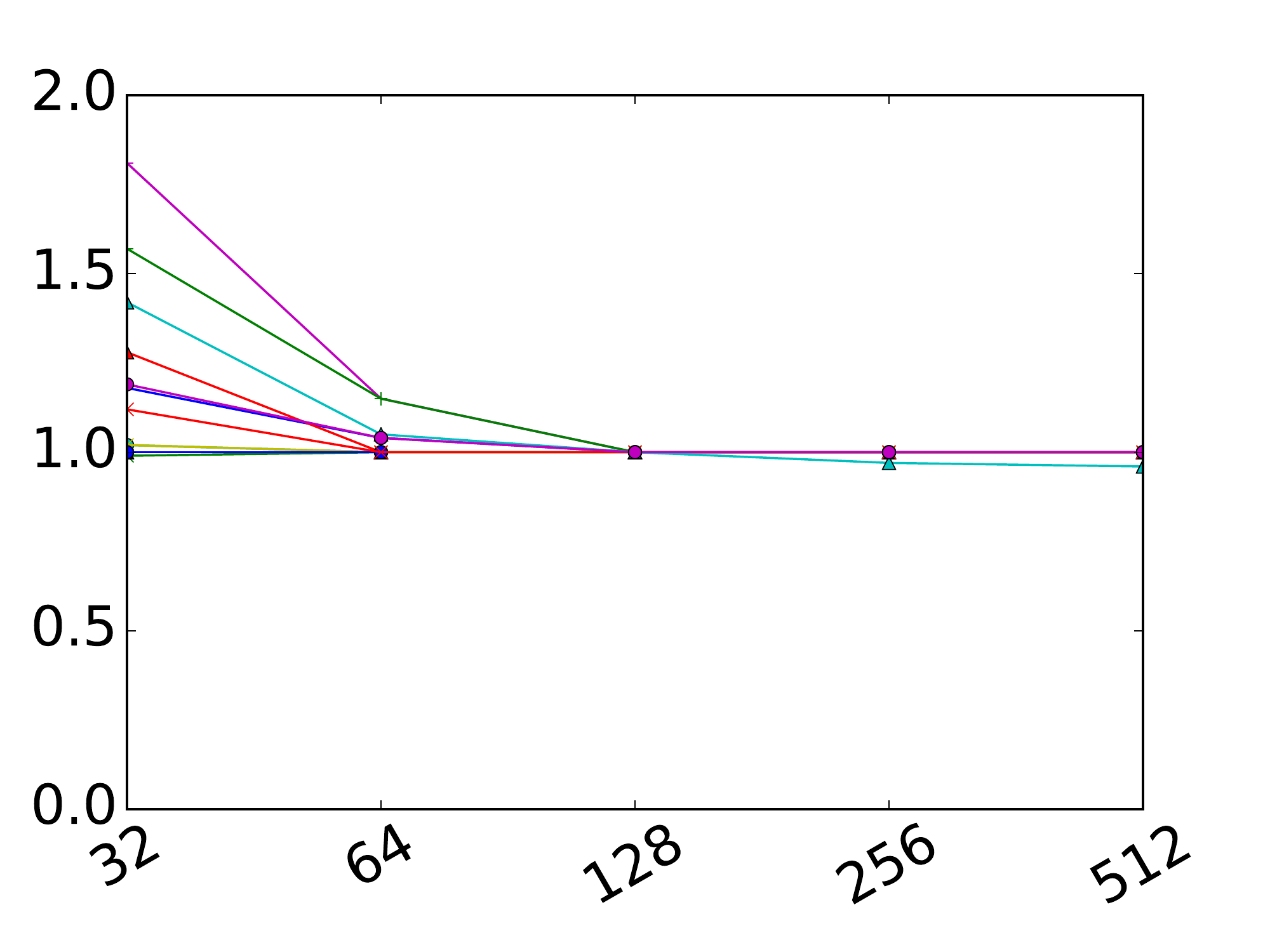}
		\caption{Number of CUDA cores}
		\label{fig:dse_cuda_cores}
	\end{subfigure}
	\begin{subfigure}[b]{0.32\textwidth}
		\includegraphics[width=\textwidth]{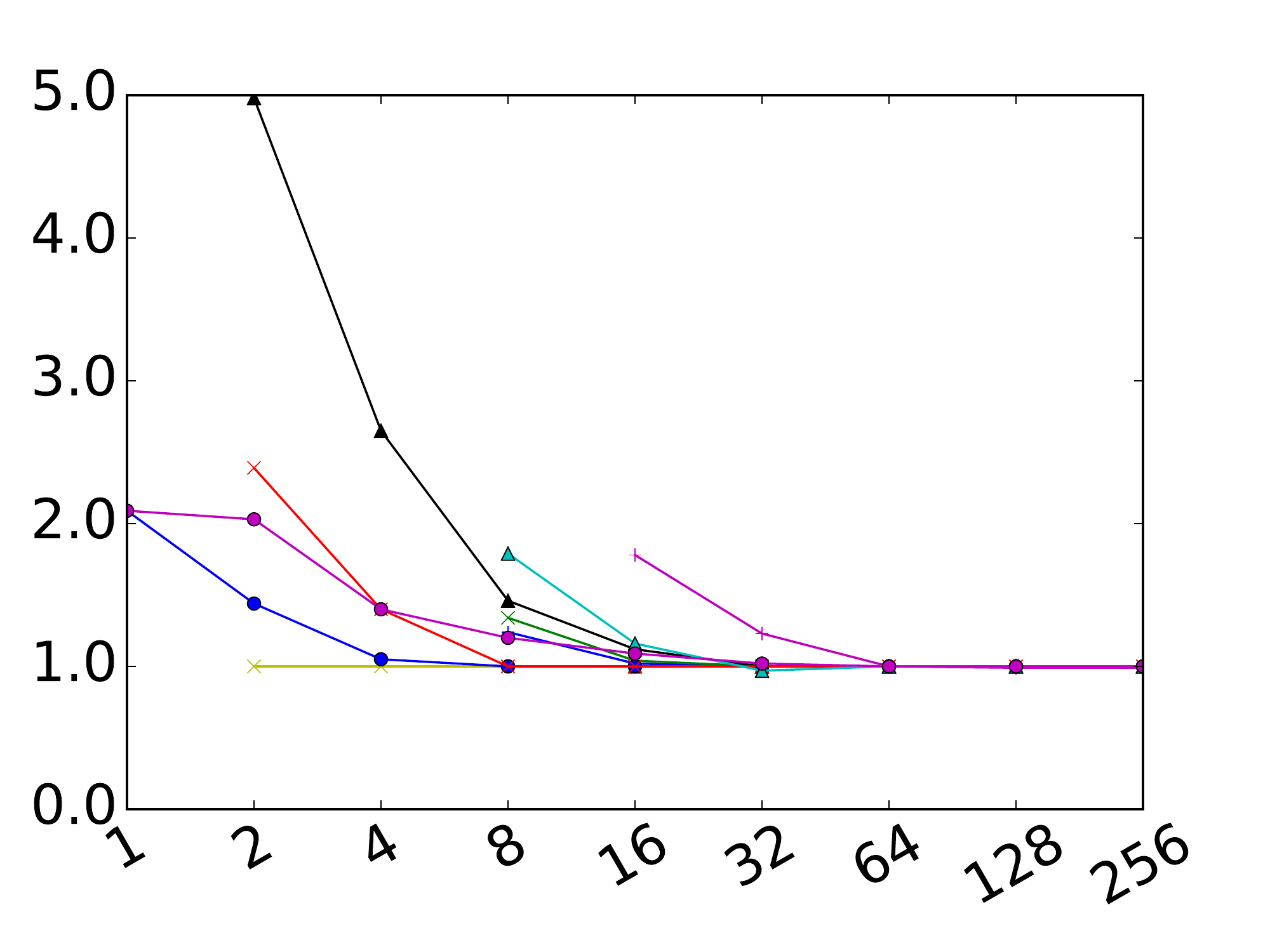}
		\caption{Register file size}
		\label{fig:dse_register_size}
	\end{subfigure}

	\begin{subfigure}[b]{0.32\textwidth}
		\includegraphics[width=\textwidth]{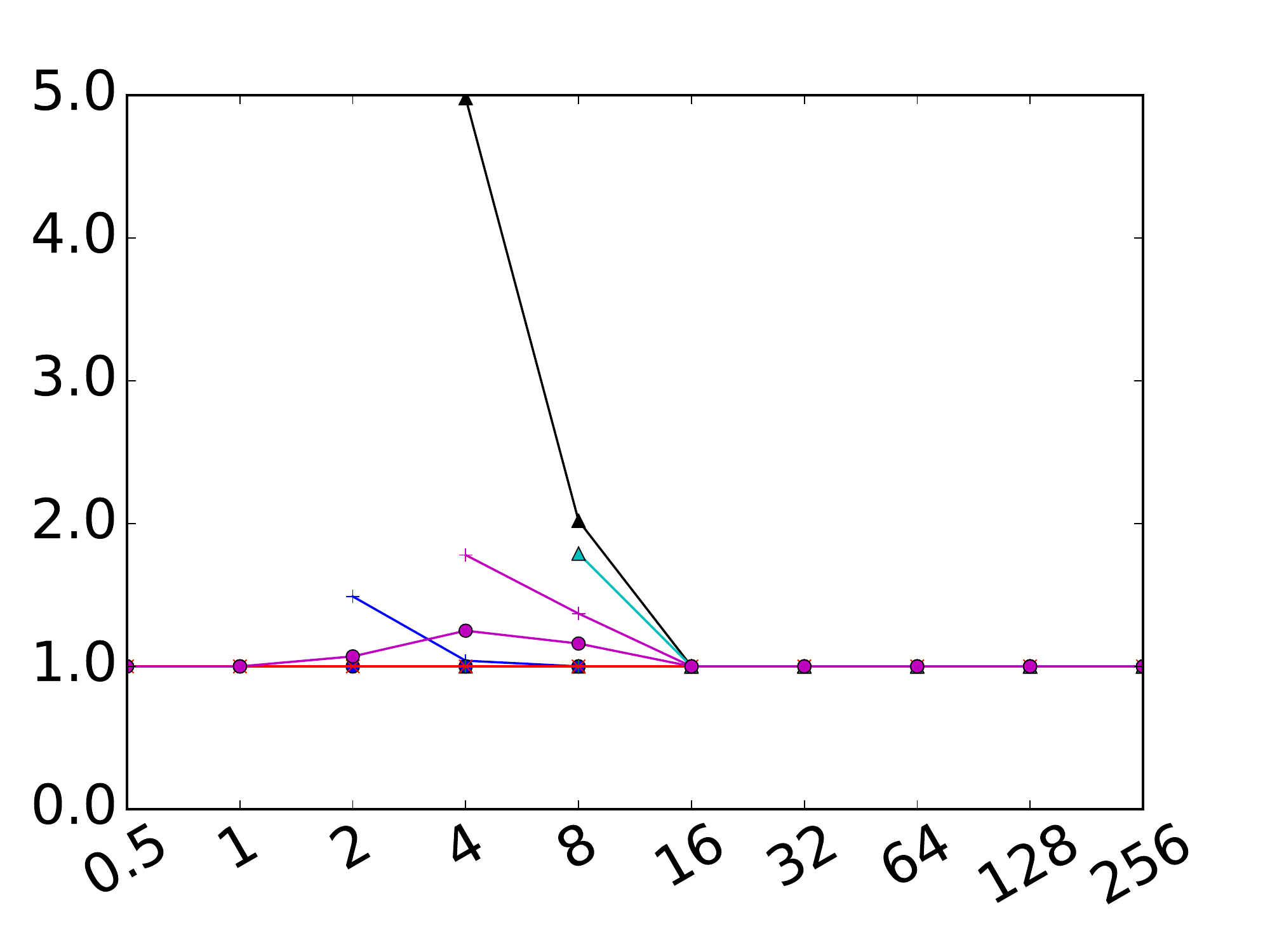}
		\caption{Shared memory size}
		\label{fig:dse_shared_mem}
	\end{subfigure}
	\begin{subfigure}[b]{0.32\textwidth}
		\includegraphics[width=\textwidth]{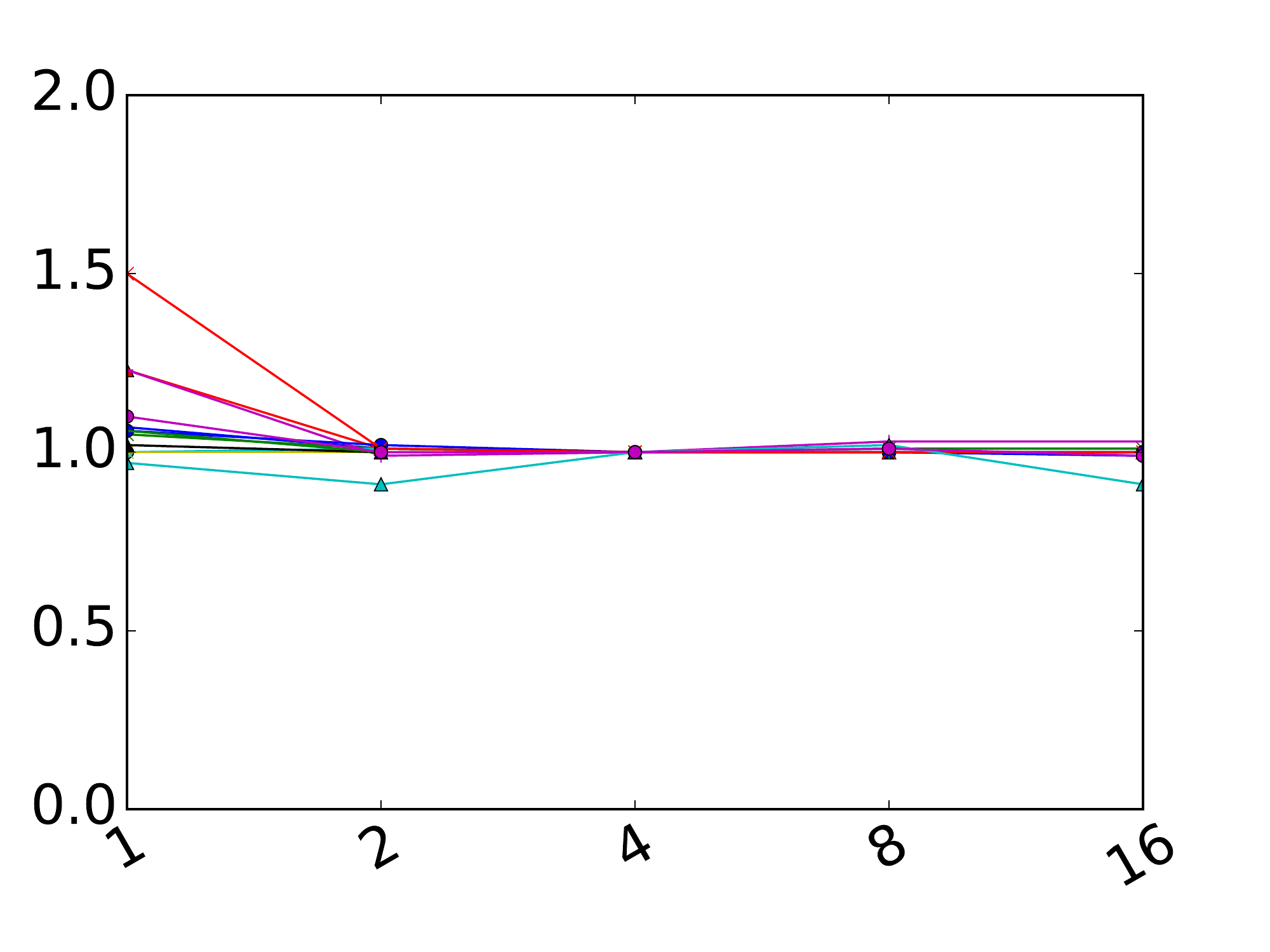}
		\caption{Number of warp schedulers}
		\label{fig:dse_warp_sch}
	\end{subfigure}
	\begin{subfigure}[b]{0.32\textwidth}
		\includegraphics[width=\textwidth]{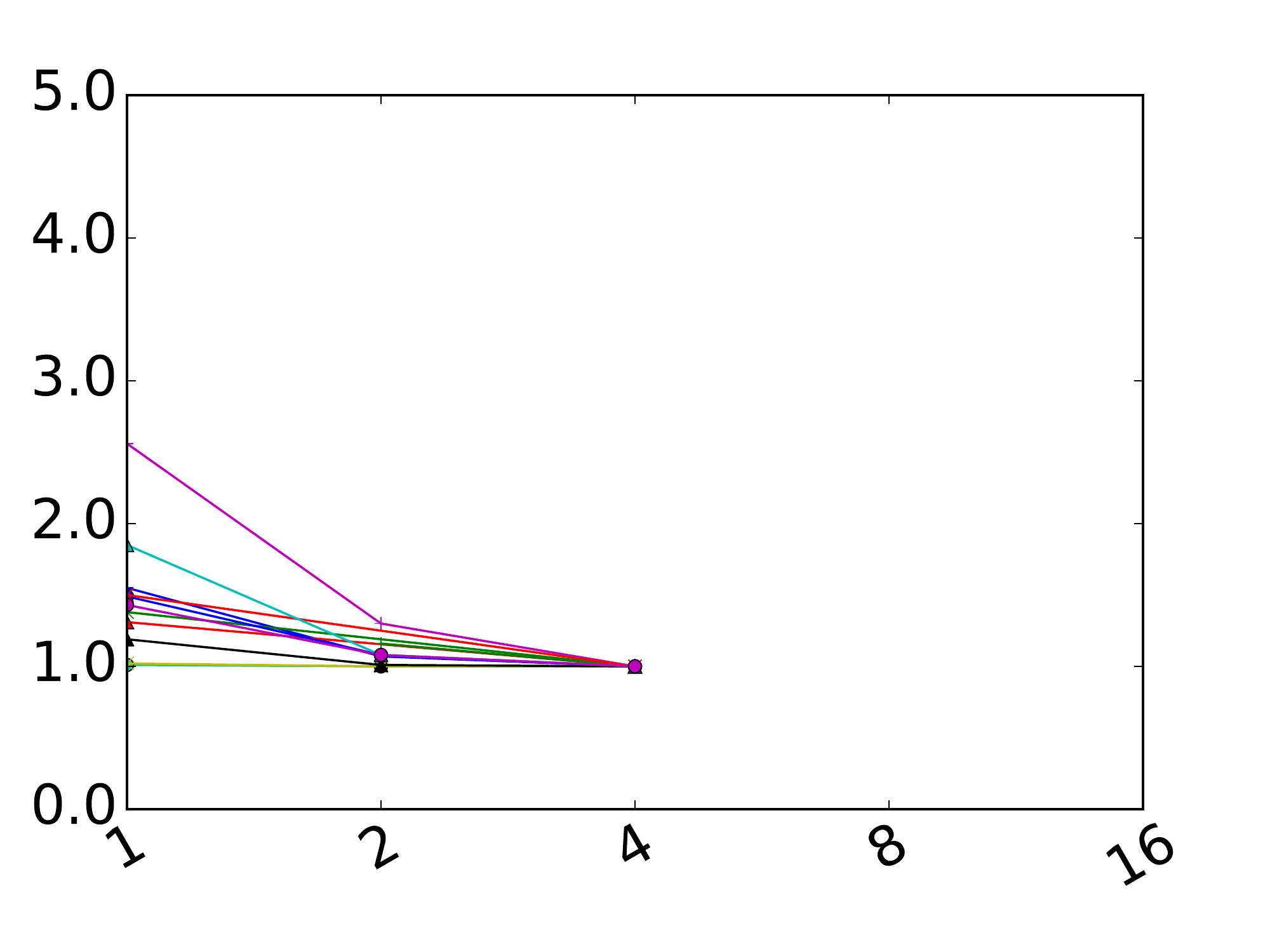}
		\caption{Number of SMB per SM}
		\label{fig:dse_smb}
	\end{subfigure}

	\begin{subfigure}[b]{0.32\textwidth}
		\includegraphics[width=\textwidth]{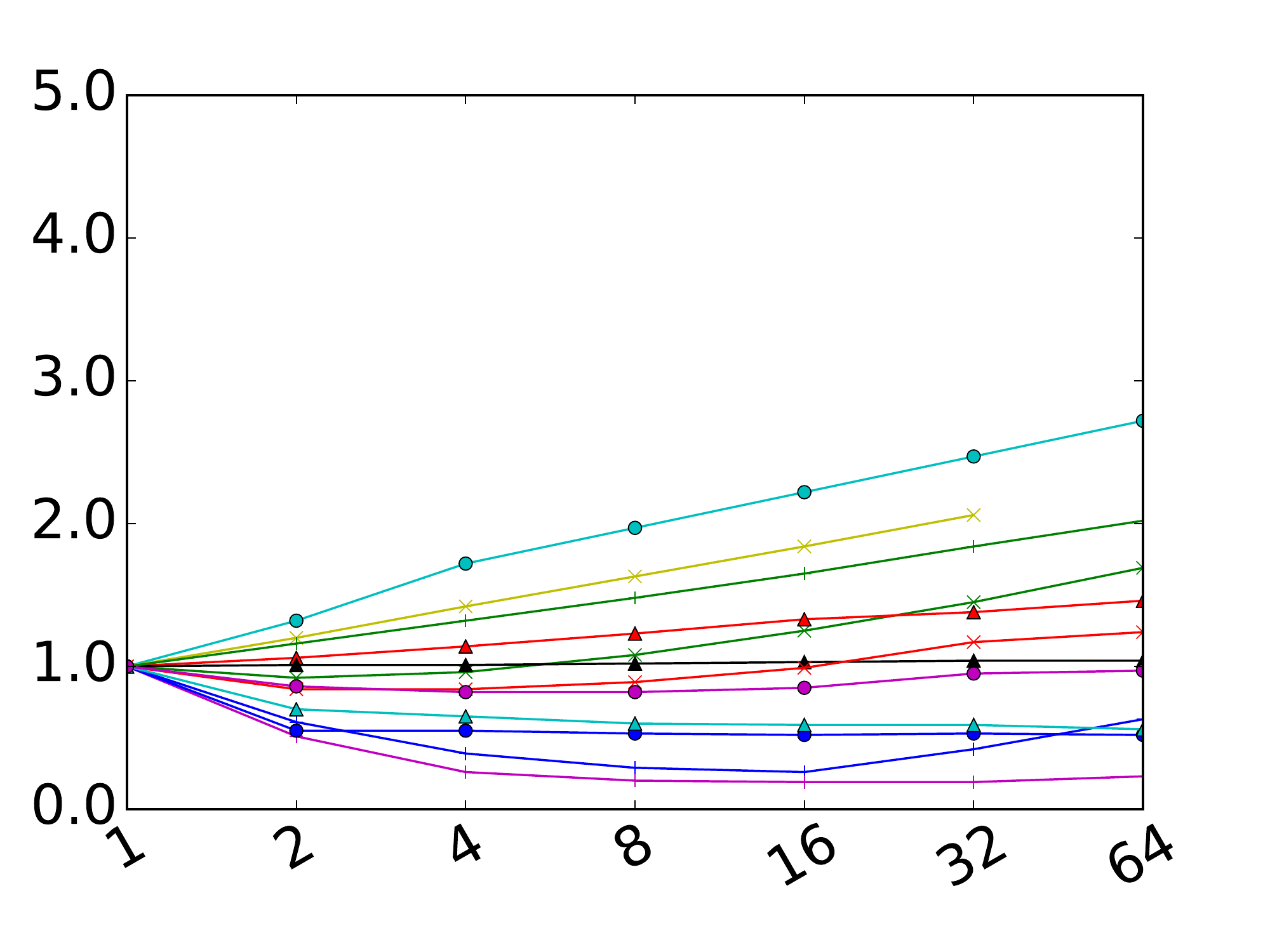}
		\caption{SM per cluster}
		\label{fig:dse_sm_clus}
	\end{subfigure}
	\begin{subfigure}[b]{0.32\textwidth}
		\includegraphics[width=\textwidth]{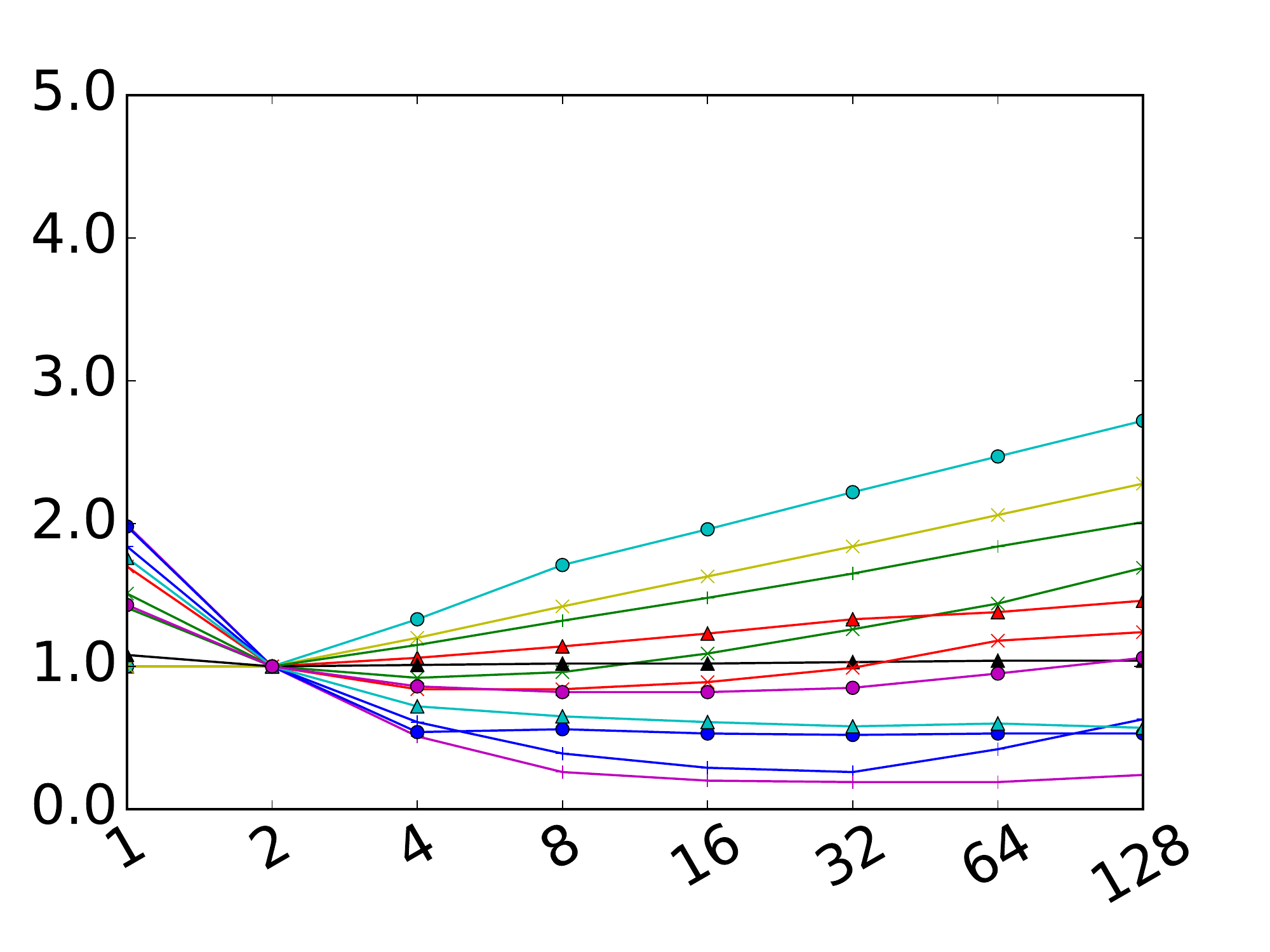}
		\caption{Number of SMs}
		\label{fig:dse_cluster}
	\end{subfigure}
	\begin{subfigure}[b]{0.32\textwidth}
		\vspace{-10em}
		\includegraphics[width=\textwidth]{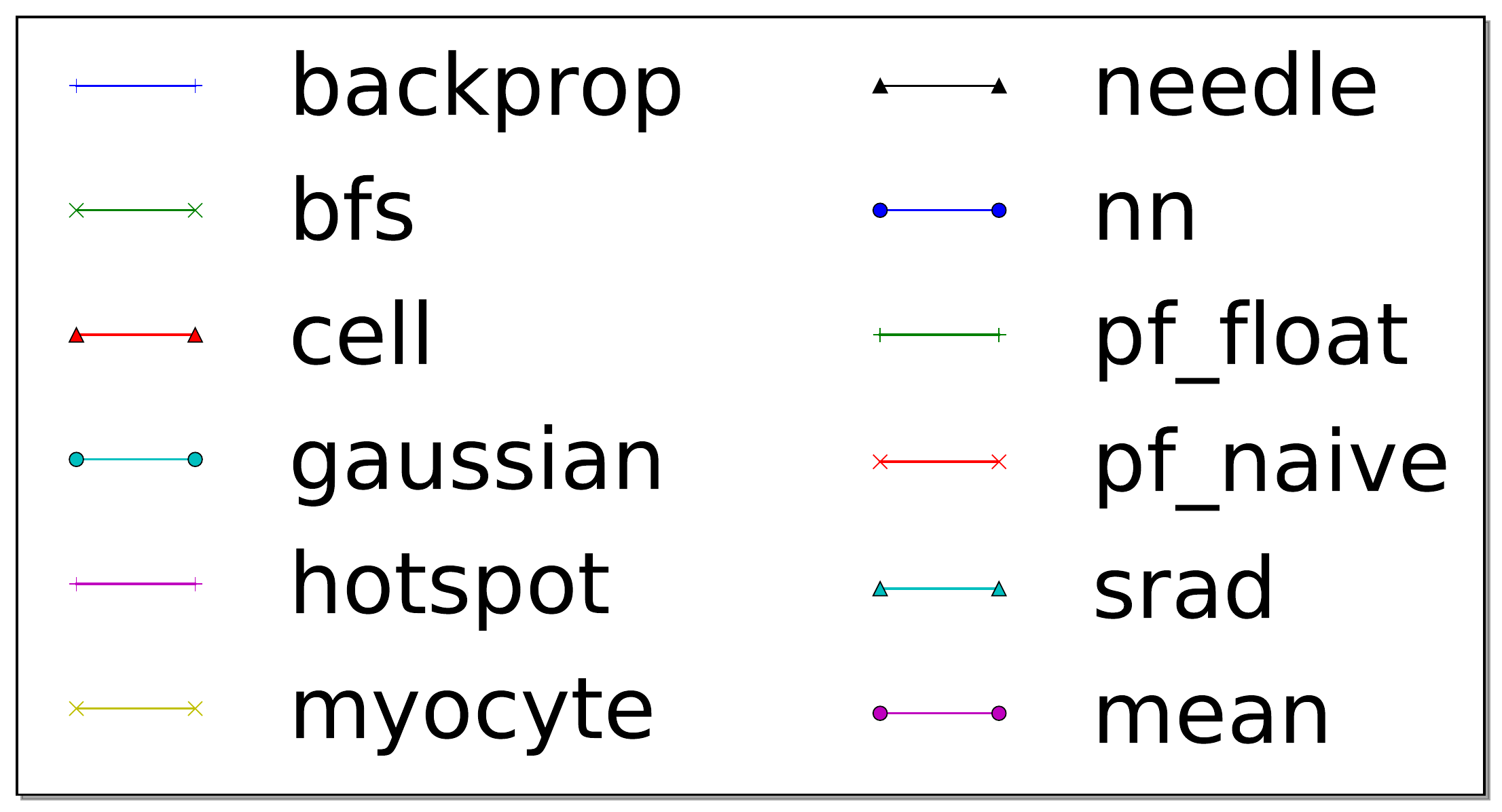}
		\caption{Legend}
		\label{fig:dse_legend}
	\end{subfigure}

	\caption{Design Space Exploration results for one parameter changes in the TX2 (all sizes are in KB).}
	\label{fig:dse-tx2}
\end{figure*}

\subsection{TX2: Changing a single parameter}

In Figure~\ref{fig:dse-tx2}, we see the results for the design space exploration when changing only one parameter. Please note that the plots have 2 different scales: 0-2 (a, b, d, e, h) and 0-5 (c, f, g, i, j, k).
Y-axis shows the slowdown in comparison with the baseline configuration.
In the first row, we show the results for the variation in L1 and L2 associativity and L2 size. Increasing the L1 size provides a small performance improvement, while reducing the size degrades the performance. When reducing the L2 cache size, we observe performance degradation, however, further increasing the L2 cache size does not provide further performance improvement.

Regarding the associativity, neither changes in L1 nor in L2 result in significant changes when increasing or decreasing it moderately (1 to 8 ways). This parameters are again tested in the next part since changing both size and associativity at the same time may have different effects that are not seen when changing them one at a time.

Figure~\ref{fig:dse-tx2} e) shows that doubling the number of CUDA cores does not increase performance, but reducing it can result in significant performance degradation, specially in compute-intensive applications like hotspot or particlefilter\_naive. In the next two Figures, f) and g), we show the effects on changing the sizes of the register file and the shared memory. Both components do not show significant differences when reducing its size by half or increasing it to double. Figure~\ref{fig:dse-tx2} h) changes the number of warp schedulers in each SM. While increasing (from 4 to 8 or 16) it does not show any performance benefits, and reducing it to just 1 incurs in significant penalty (10\% more execution time), reducing it by half (from 4 to 2) results in the same average performance, with almost no variability in the different benchmarks.

\begin{figure*}[t!]
	\centering

	\begin{subfigure}[b]{0.32\textwidth}
		\includegraphics[width=\textwidth]{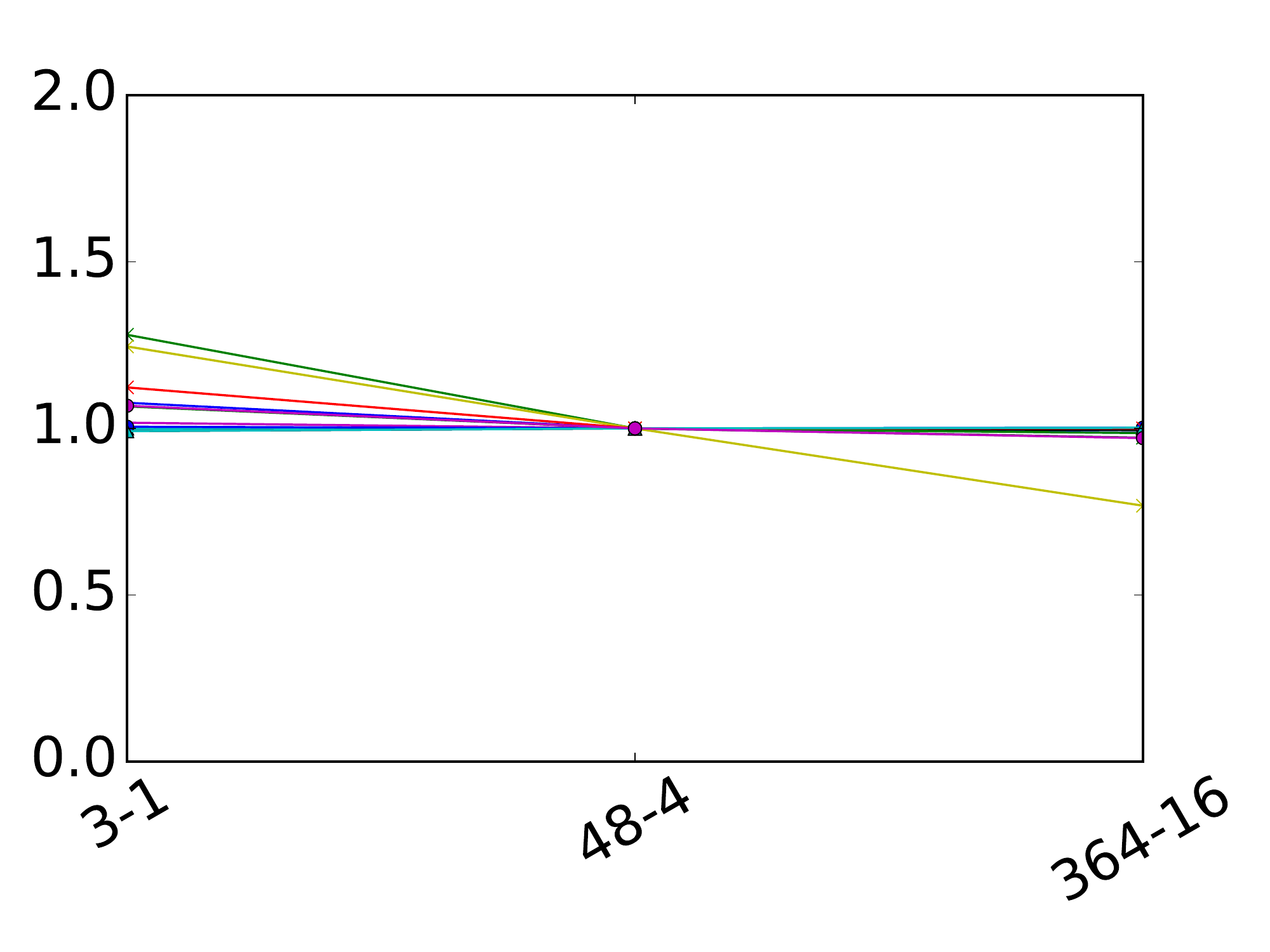}
		\caption{L1 size and associativity}
		\label{fig:dsem_l1_sz_as}
	\end{subfigure}
	\begin{subfigure}[b]{0.32\textwidth}
		\includegraphics[width=\textwidth]{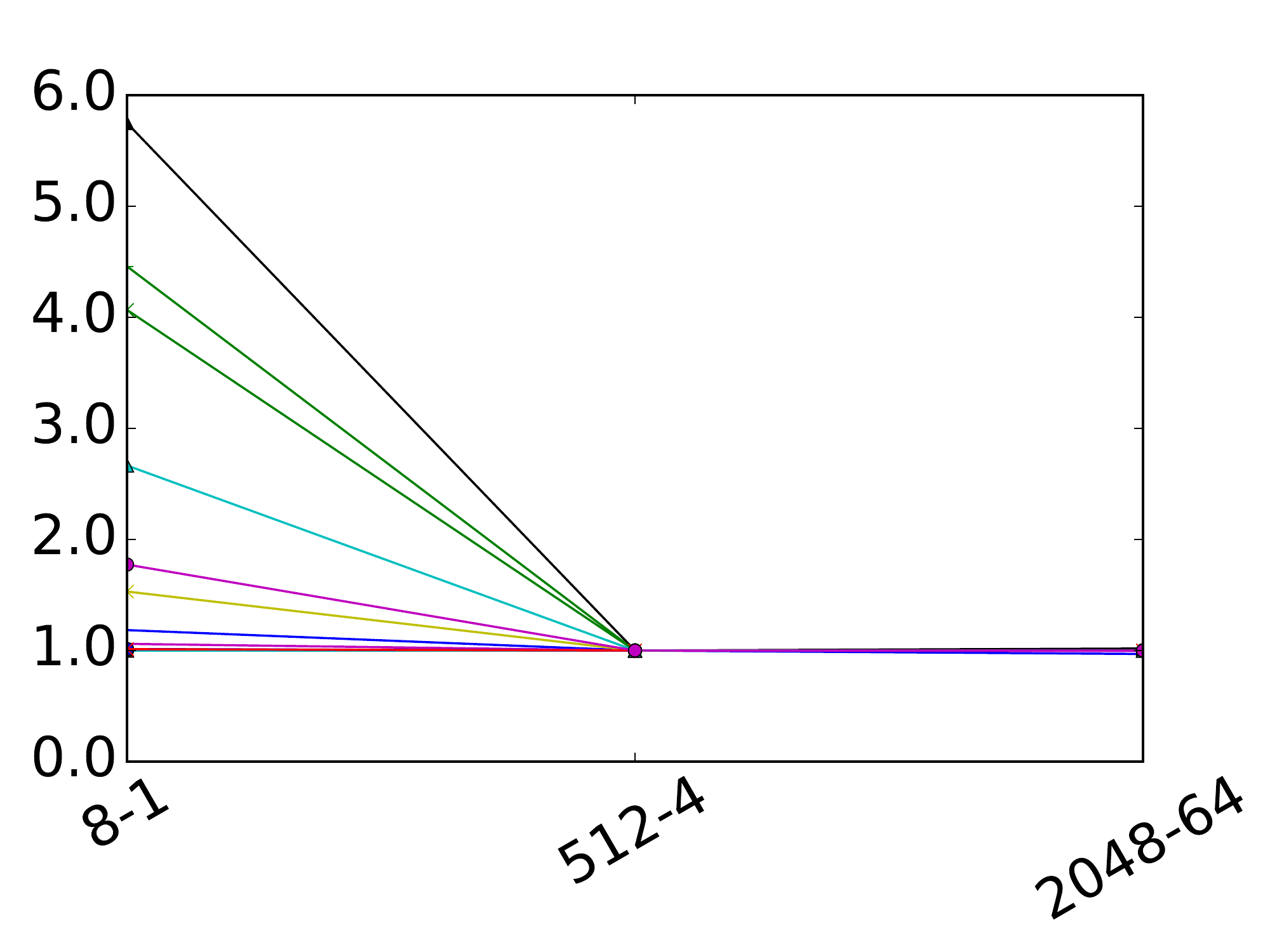}
		\caption{L2 size and associativity}
		\label{fig:dsem_l2_sz_as}
	\end{subfigure}
	\begin{subfigure}[b]{0.32\textwidth}
		\includegraphics[width=\textwidth]{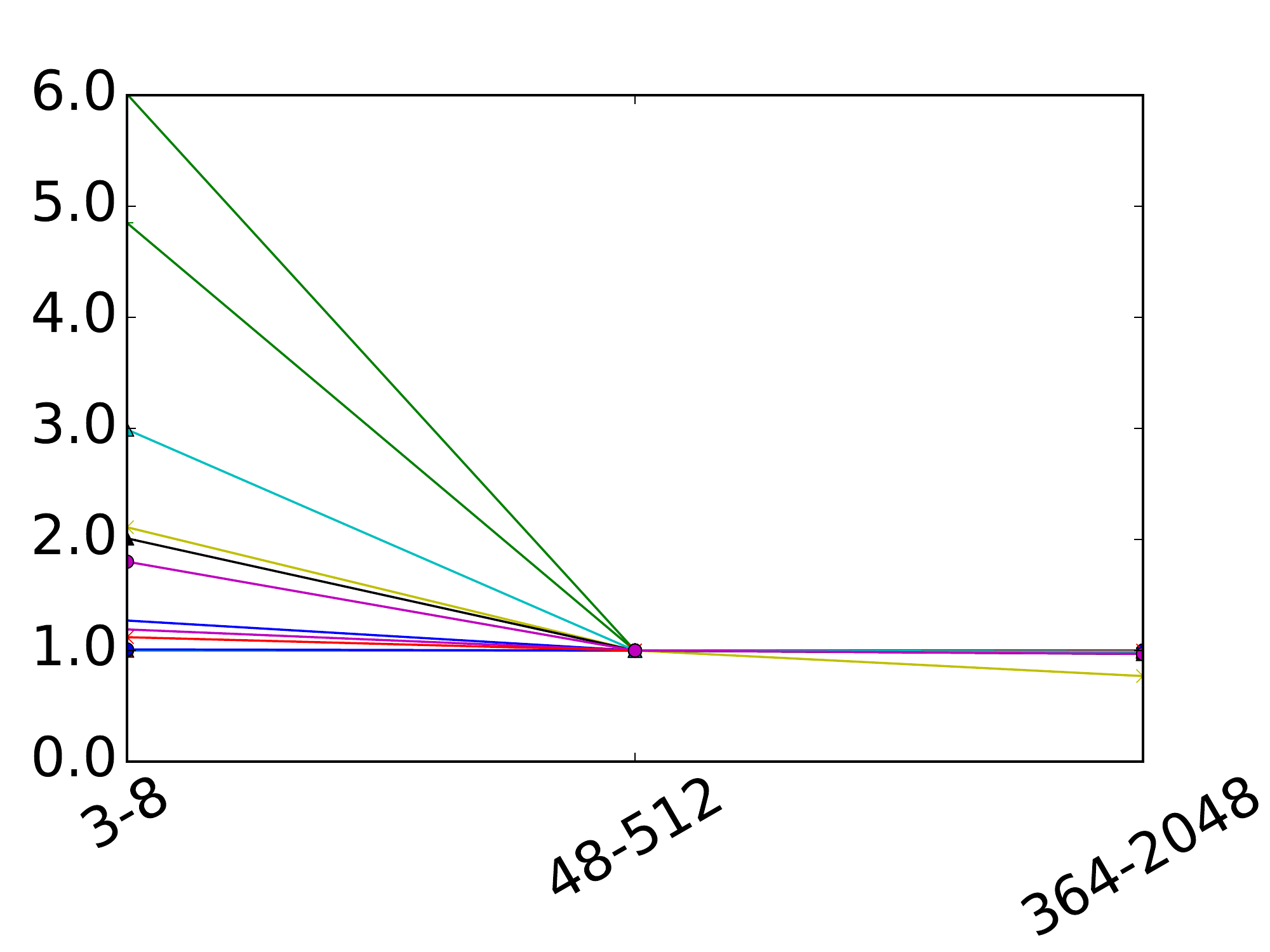}
		\caption{L1 and L2 size}
		\label{fig:dsem_l1_l2_sz}
	\end{subfigure}

	\begin{subfigure}[b]{0.32\textwidth}
		\includegraphics[width=\textwidth]{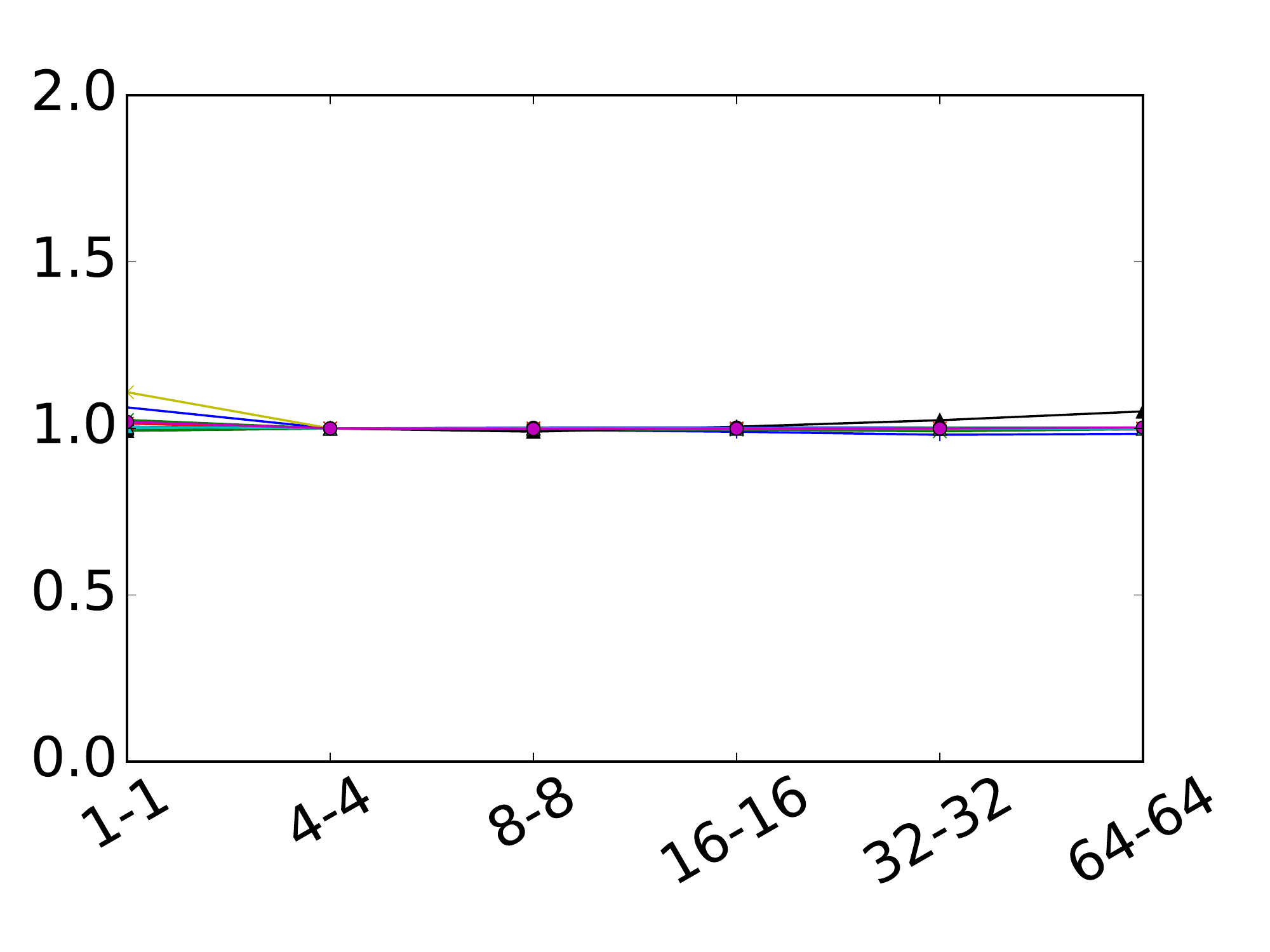}
		\caption{L1 and L2 associativity}
		\label{fig:dsem_l1_l2_as}
	\end{subfigure}
	\begin{subfigure}[b]{0.32\textwidth}
		\includegraphics[width=\textwidth]{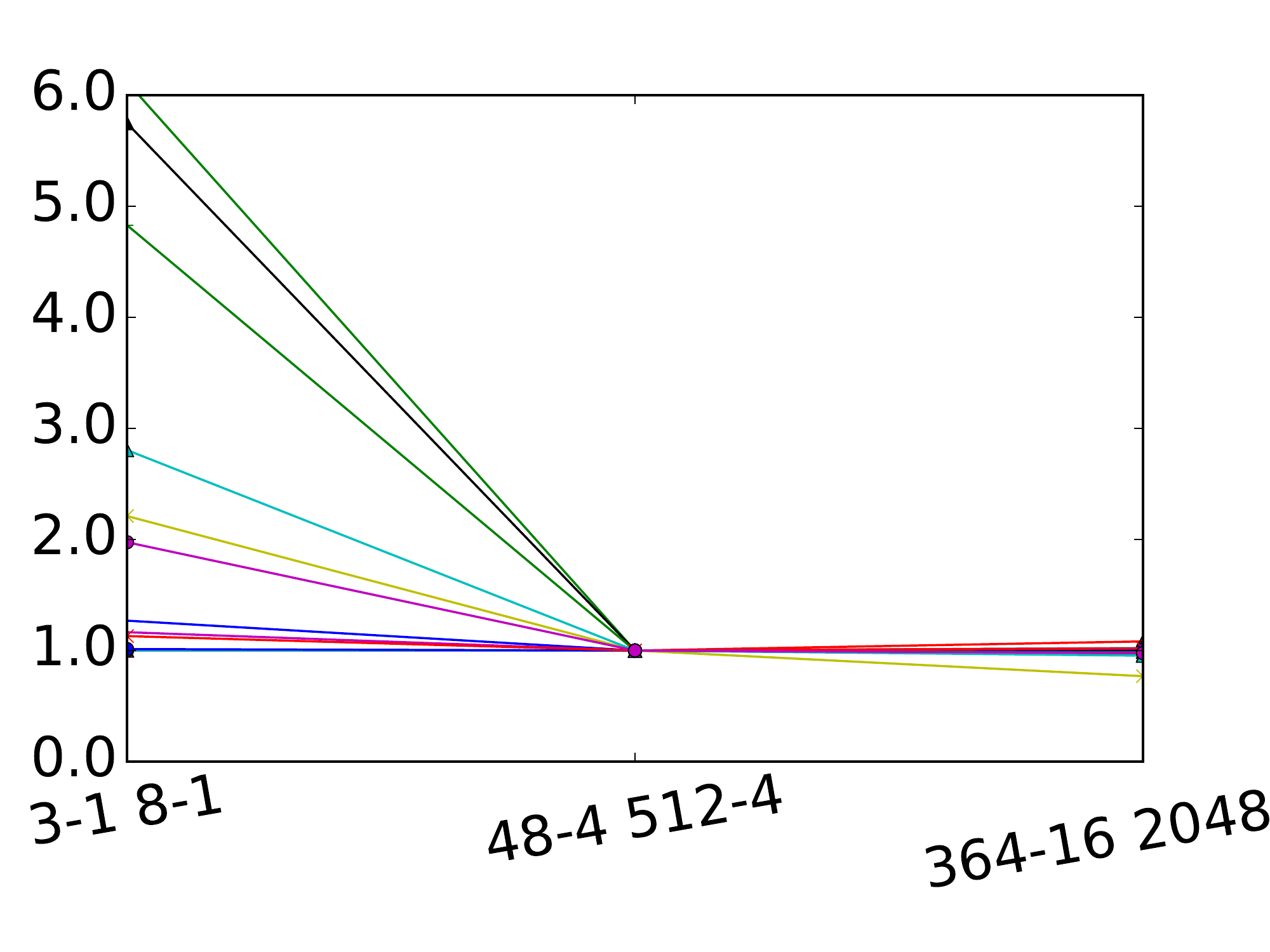}
		\caption{L1 and L2 size and associativity}
		\label{fig:dsem_l1_l2_sz_as}
	\end{subfigure}
	\begin{subfigure}[b]{0.32\textwidth}
		\includegraphics[width=\textwidth]{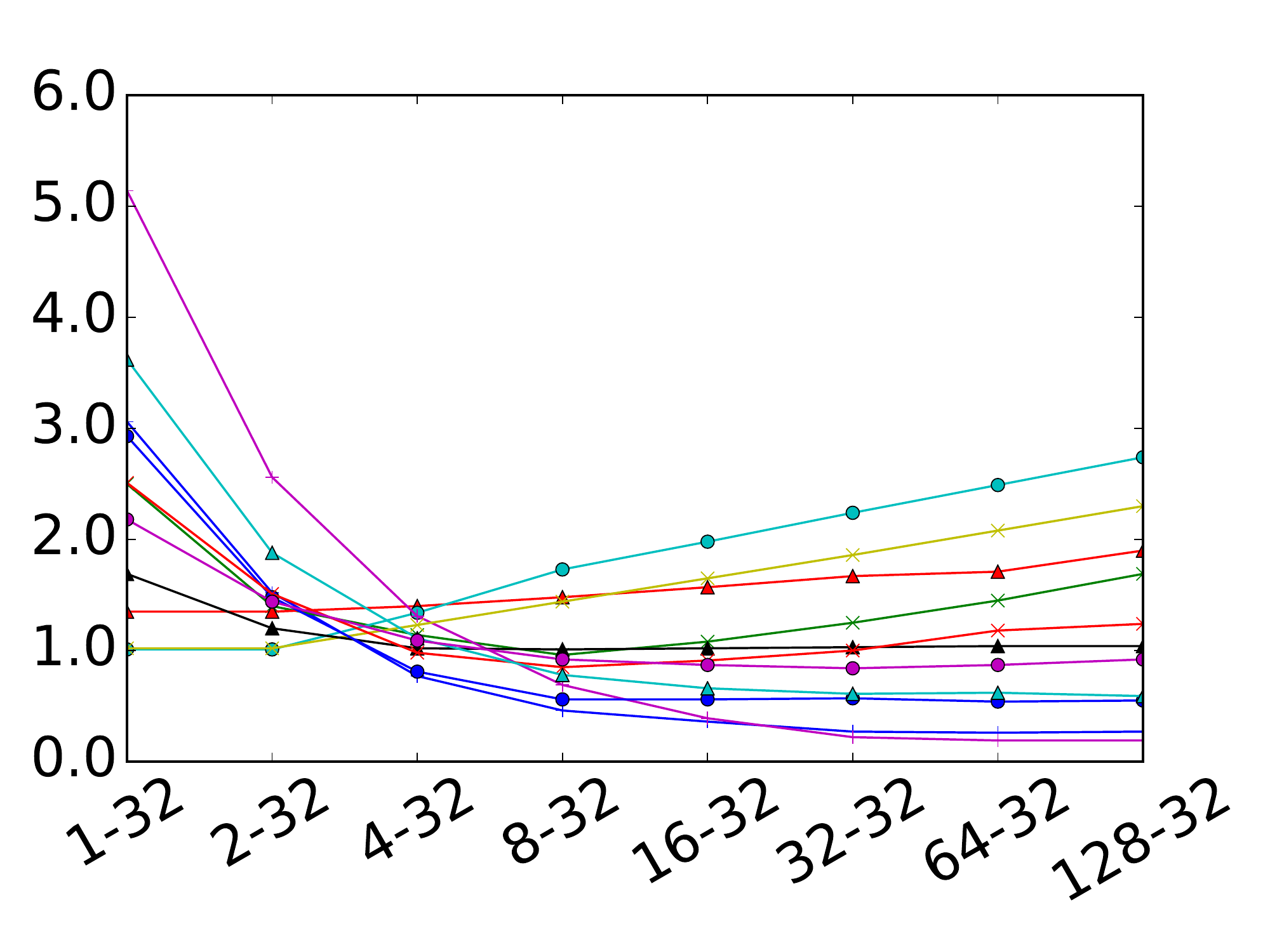}
		\caption{SM with 32 CUDA cores}
		\label{fig:dsem_sm_32}
	\end{subfigure}

	\caption{Design Space Exploration results for several parameter changes in the TX2 (all sizes in KB). The legend is the same used in Figure~\ref{fig:dse-tx2}.}
	\label{fig:dsem}
\end{figure*}

In Figures~\ref{fig:dse-tx2} i),~\ref{fig:dse-tx2} j), and ~\ref{fig:dse-tx2} k), we show the results for the number of SMB per SM, the number of SM per cluster, and the number of clusters with just 1 SM, respectively. Since the simulator does not support more than 4 SMB per SM (the standard in Pascal), we just focus on reducing it to 1 and 2. We observe that this reduction would incur in significant performance penalties. The last two (SM per cluster and number of clusters) show a similar trend, since in the end both of them are increasing the total amount of SMs. We observe that increasing the total number of SMs has positive effects in performance, but just up to a certain point. The increase in SMs has to be tailored to the application implementation and behavior, so some applications may not be able to use all these SMs while others could if they were implemented with more granularity. These two components are the ones that vary most depending on the benchmark, with several improving performance and others decreasing it.

\subsection{TX2: Changing two or more parameters}

Some parameters that changed on their own have a specific impact on performance, can behave differently if changed together with other parameters. This is because how one parameter performs depends also on how other parameters perform. This is obvious in caches. Changing the associativity of a small cache may have bigger impact on performance than changing the associativity of a big cache, since sets may be big enough anyway to notice any degradation.

The first two parameters that we have changed together are cache size and associativity, both for L1 (Figure~\ref{fig:dsem_l1_sz_as}) and L2 (Figure~\ref{fig:dsem_l2_sz_as}). For both we have changed the associativity (first element in the x-axis) and the size (second element) together. Although for the L1 we do not see significant performance changes, for the L2 we see that reducing the size and associativity to small enough numbers we have significant performance degradation. This degradation is bigger than the one observed in the previous experiments when only changing the associativity or the size separately.

In the next experiments, we change the size of both caches at the same time (Figure~\ref{fig:dsem_l1_l2_sz}), and we see similar results to the previous experiment, with a lot of degradation when the L2 size is small. We also changed both L1 and L2 associativities at the same time (Figure~\ref{fig:dsem_l1_l2_as}) and as in previous experiments we see that it has no noticeable effect. In the last experiment with multiple cache components, we change both L1 and L2 sizes and associativities. The results are similar to b) and c), with big performance penalties in the small setup and no improvement in the big one.

Finally, in Figure~\ref{fig:dsem_sm_32}, we change both the number of SM as well as the number of CUDA cores per SM, changed from 128 to just 32. Comparing with the previous experiment, Figure~\ref{fig:dse_cluster}, we see that the trends are similar, with small variations depending on the benchmark.

{\color{black}

\begin{figure*}[h!]
	\centering
	
	\begin{subfigure}[b]{0.32\textwidth}
		\includegraphics[width=\textwidth]{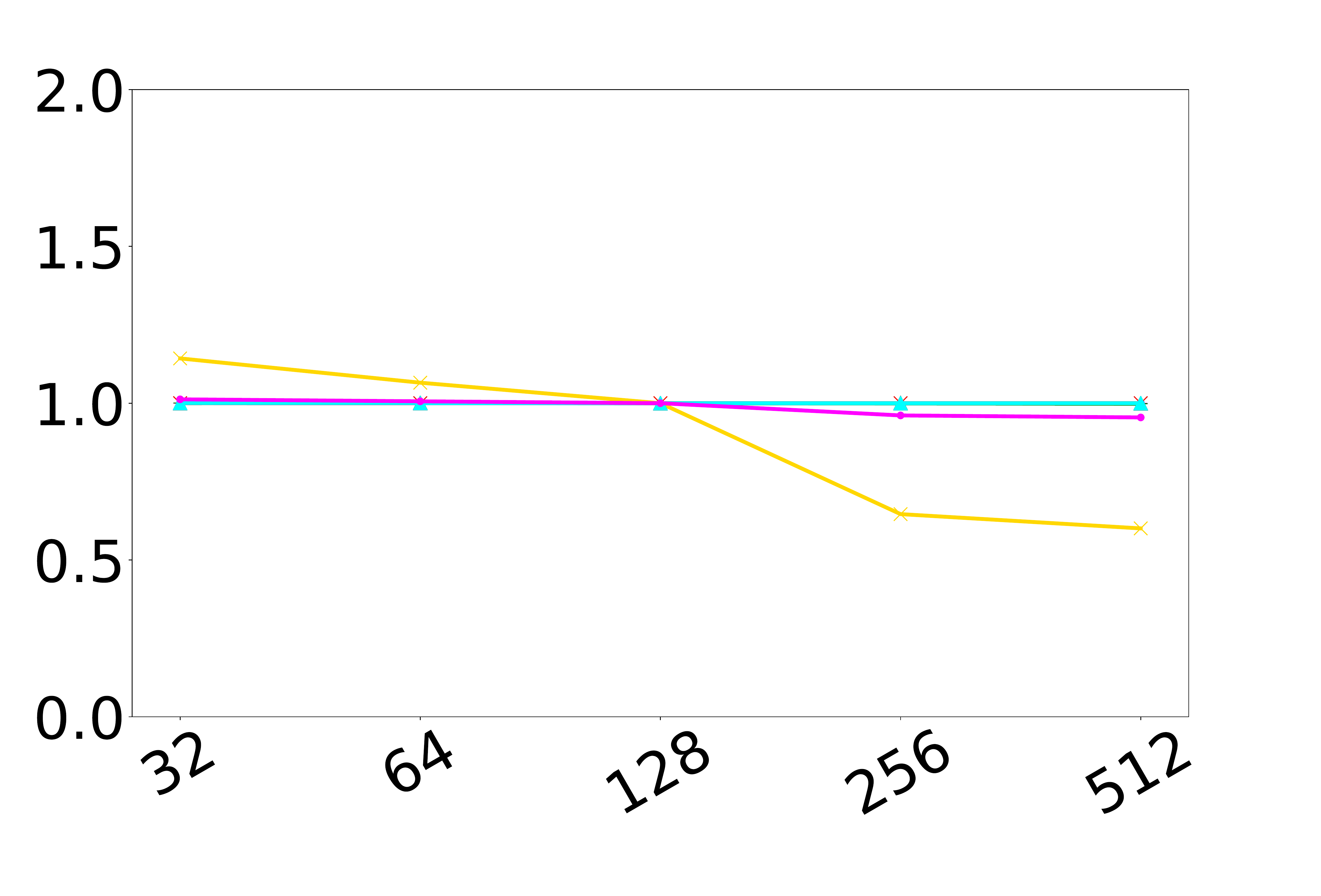}
		\caption{L1 size}
		\label{fig:dse_l1size-xavier}
	\end{subfigure}
	\begin{subfigure}[b]{0.32\textwidth}
		\includegraphics[width=\textwidth]{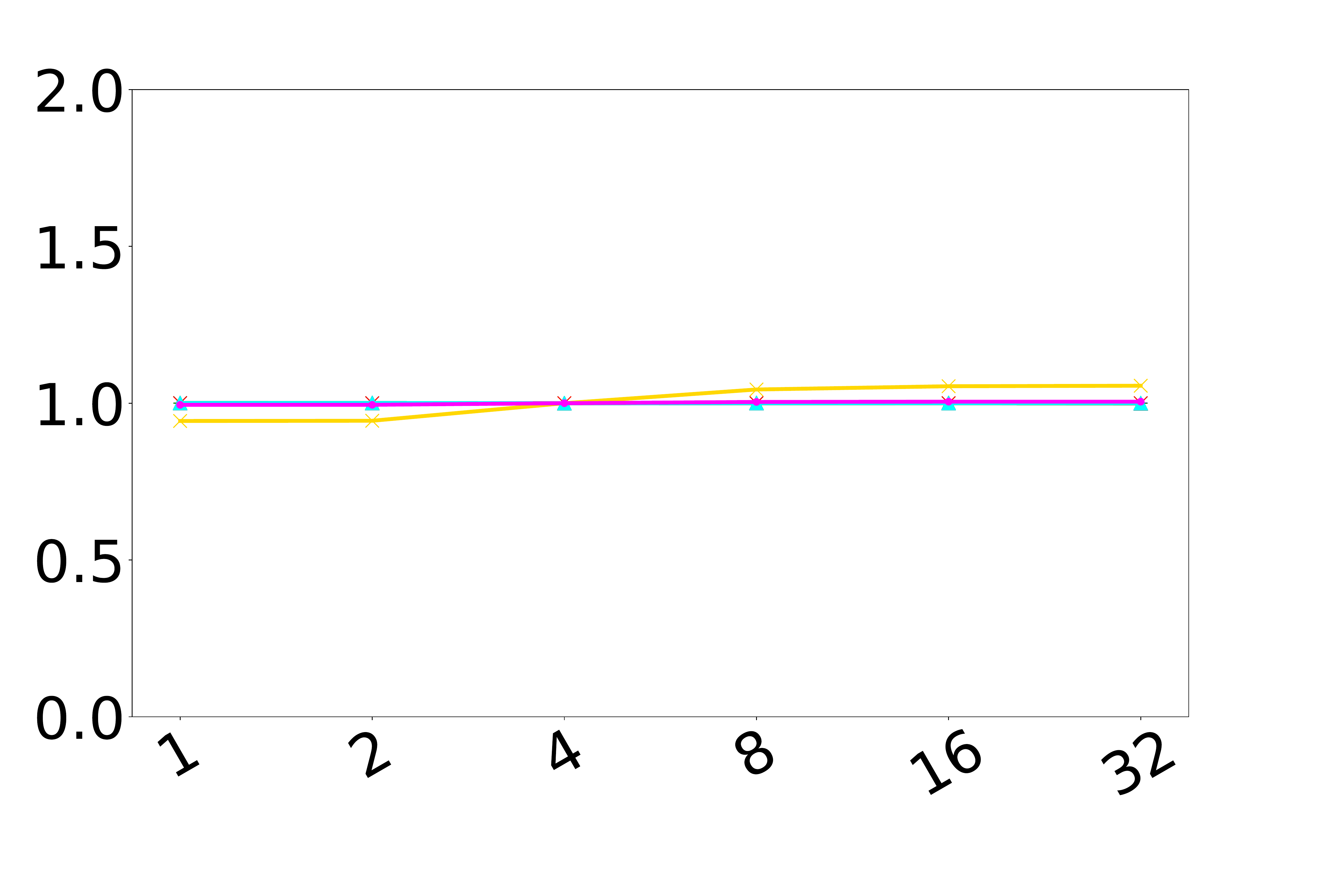}
		\caption{L1 associativity}
		\label{fig:dse_l1assoc-xavier}
	\end{subfigure}
	\begin{subfigure}[b]{0.32\textwidth}
		\includegraphics[width=\textwidth]{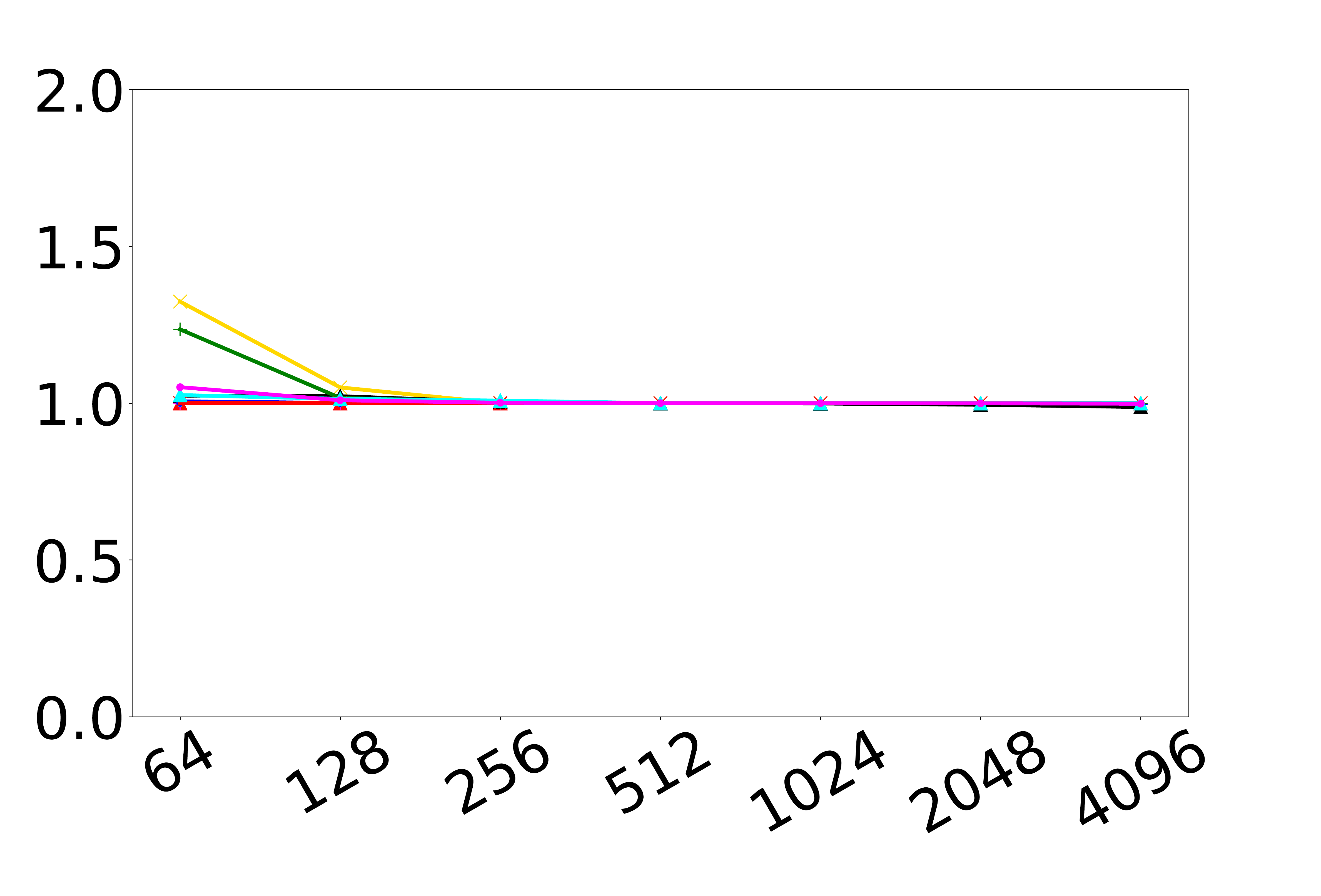}
		\caption{L2 size}
		\label{fig:dse_l2size-xavier}
	\end{subfigure}
	
	\begin{subfigure}[b]{0.32\textwidth}
		\includegraphics[width=\textwidth]{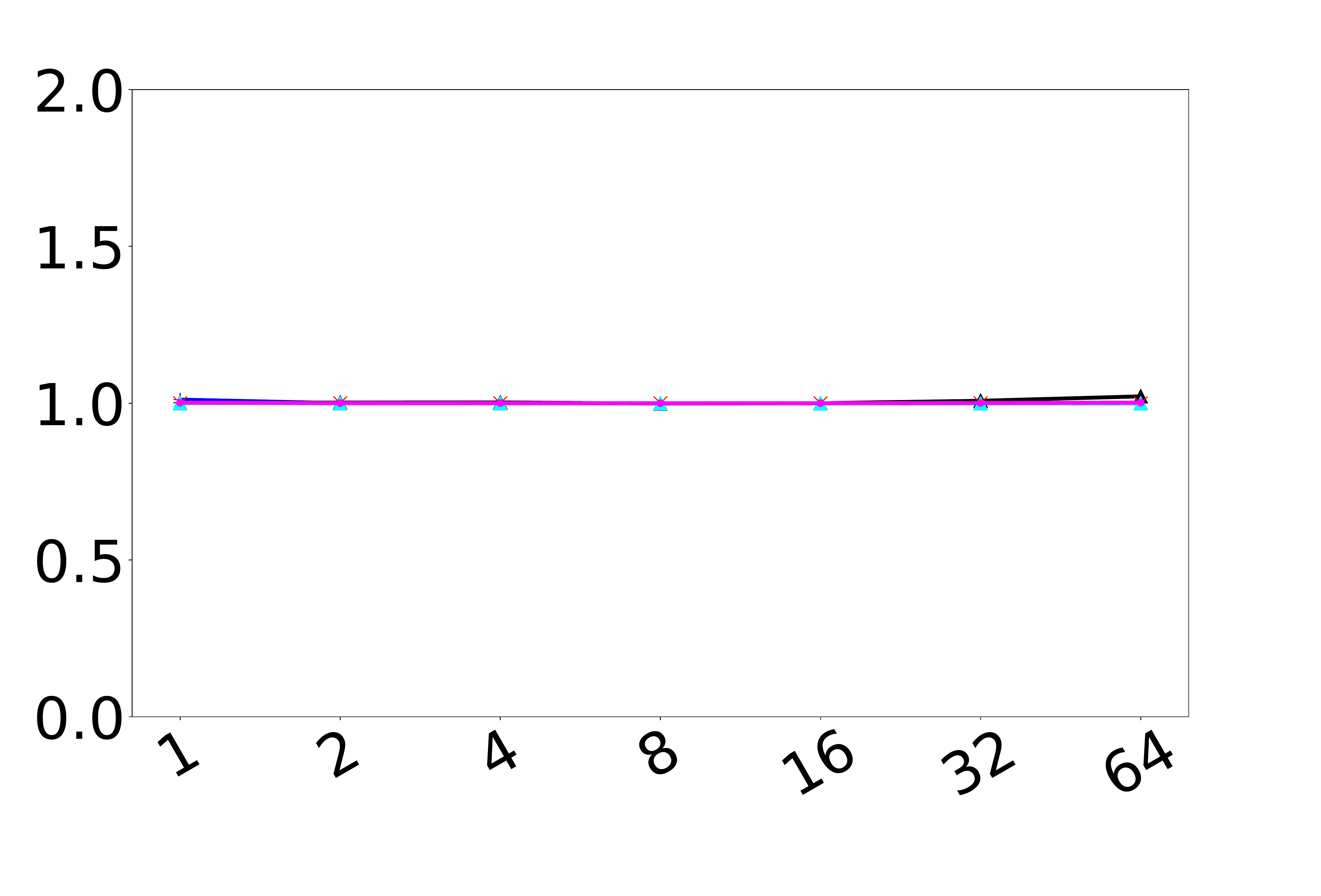}
		\caption{L2 associativity}
		\label{fig:dse_l2assoc-xavier}
	\end{subfigure}
	\begin{subfigure}[b]{0.32\textwidth}
		\includegraphics[width=\textwidth]{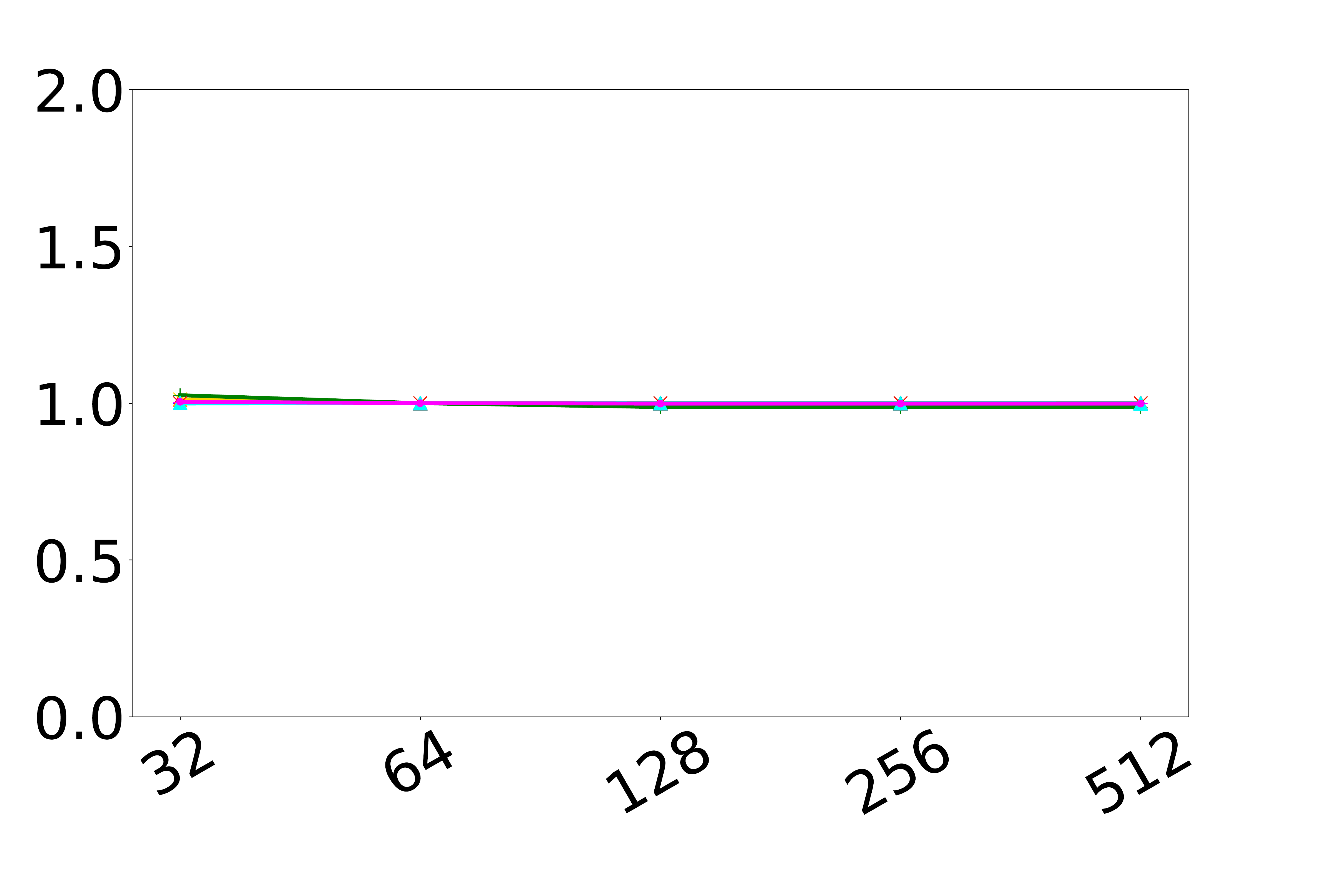}
		\caption{Number of CUDA cores}
		\label{fig:dse_cuda_cores-xavier}
	\end{subfigure}
	\begin{subfigure}[b]{0.32\textwidth}
		\includegraphics[width=\textwidth]{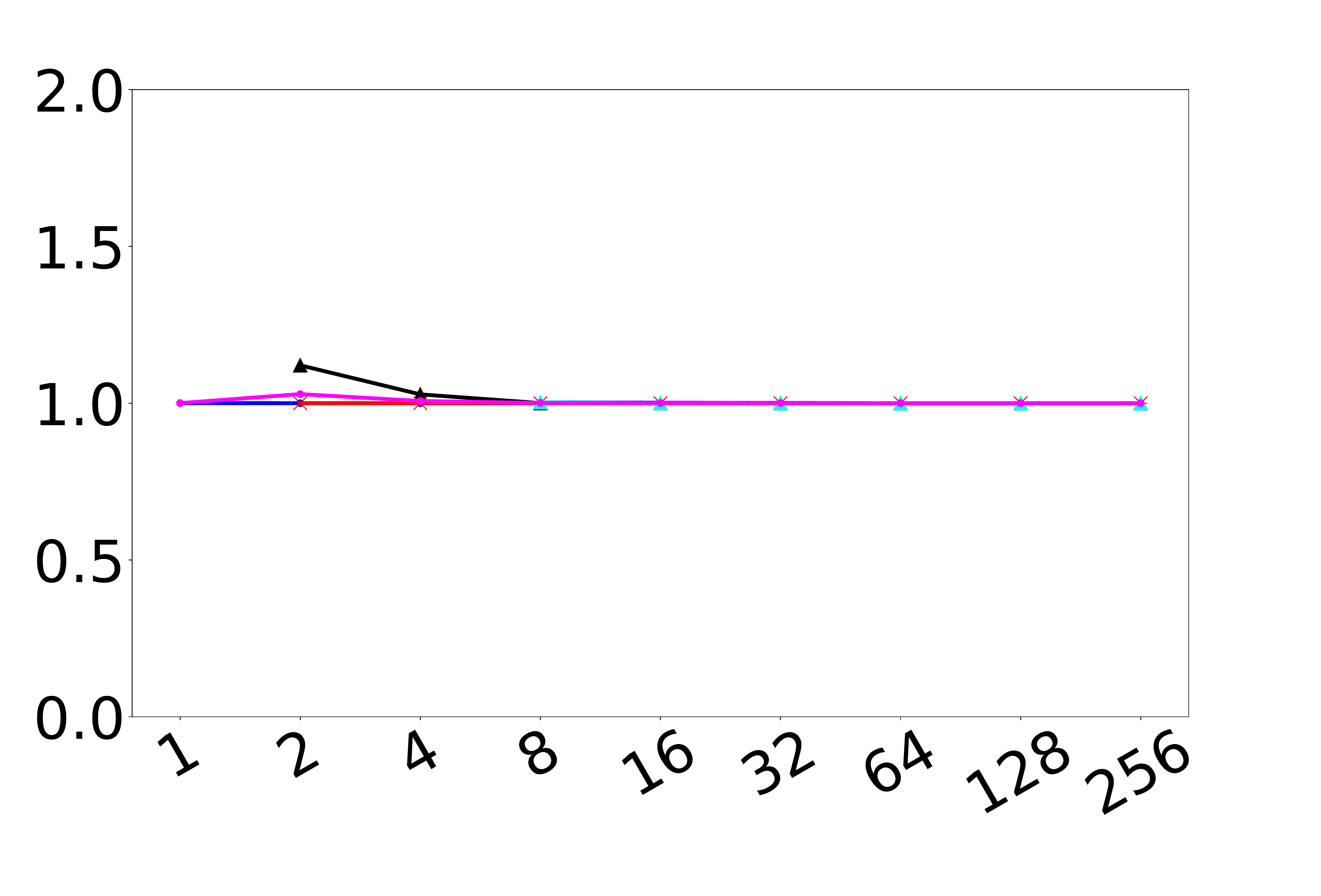}
		\caption{Register file size}
		\label{fig:dse_register_size-xavier}
	\end{subfigure}
	
	\begin{subfigure}[b]{0.32\textwidth}
		\includegraphics[width=\textwidth]{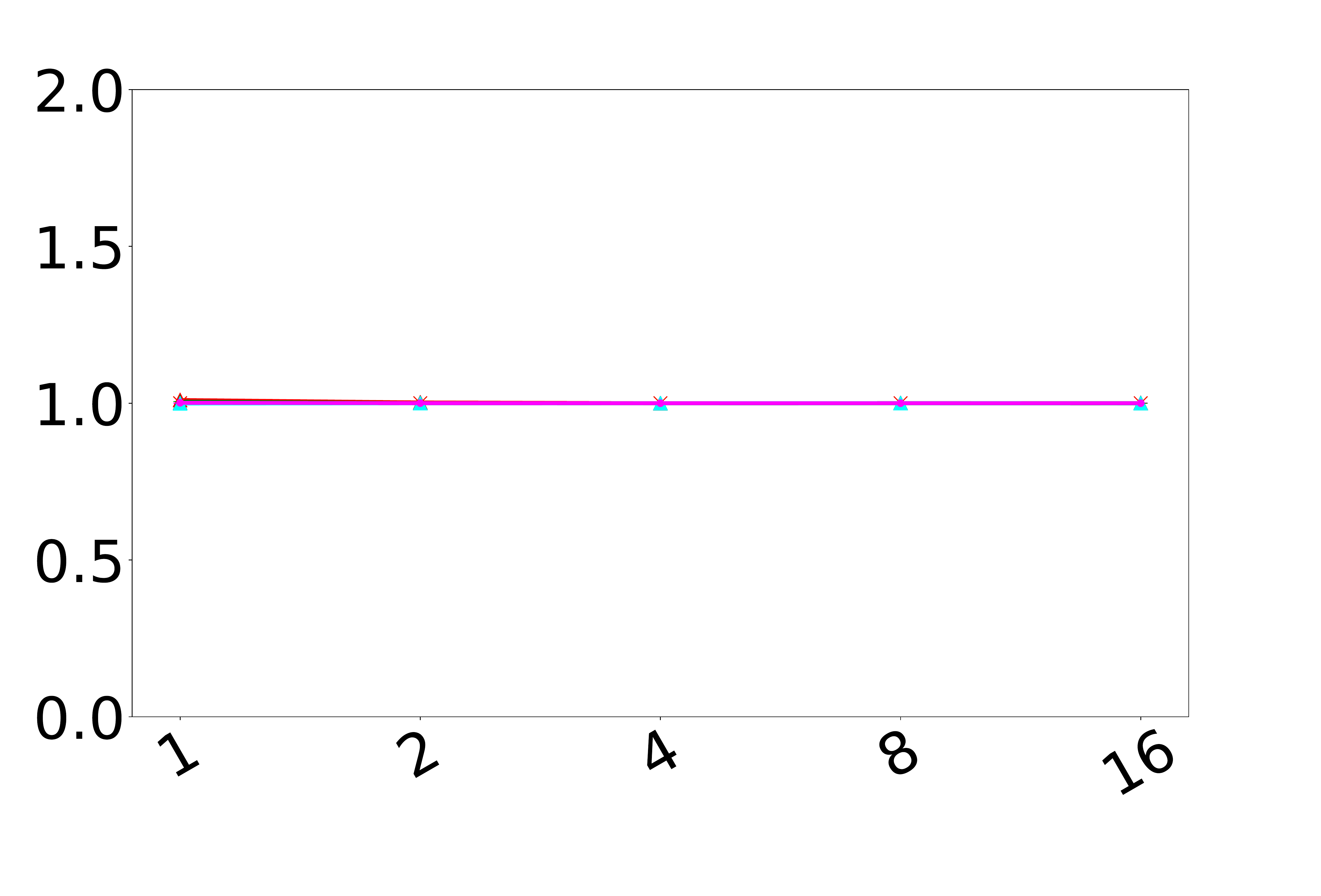}
		\caption{Number of warp schedulers}
		\label{fig:dse_warp_sch-xavier}
	\end{subfigure}
	\begin{subfigure}[b]{0.32\textwidth}
		\includegraphics[width=\textwidth]{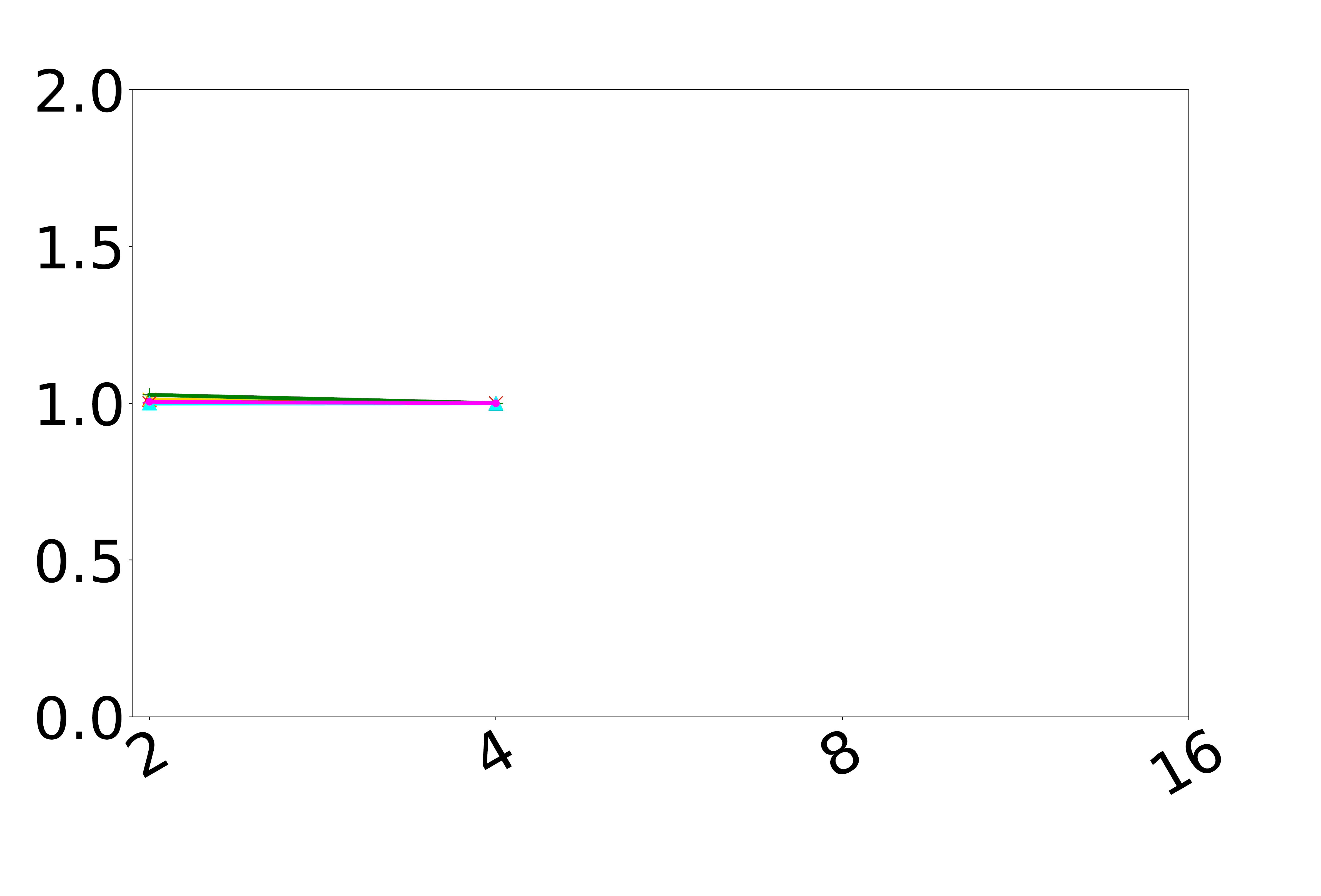}
		\caption{Number of SMB per SM}
		\label{fig:dse_smb-xavier}
	\end{subfigure}
	
	\begin{subfigure}[b]{0.32\textwidth}
		\includegraphics[width=\textwidth]{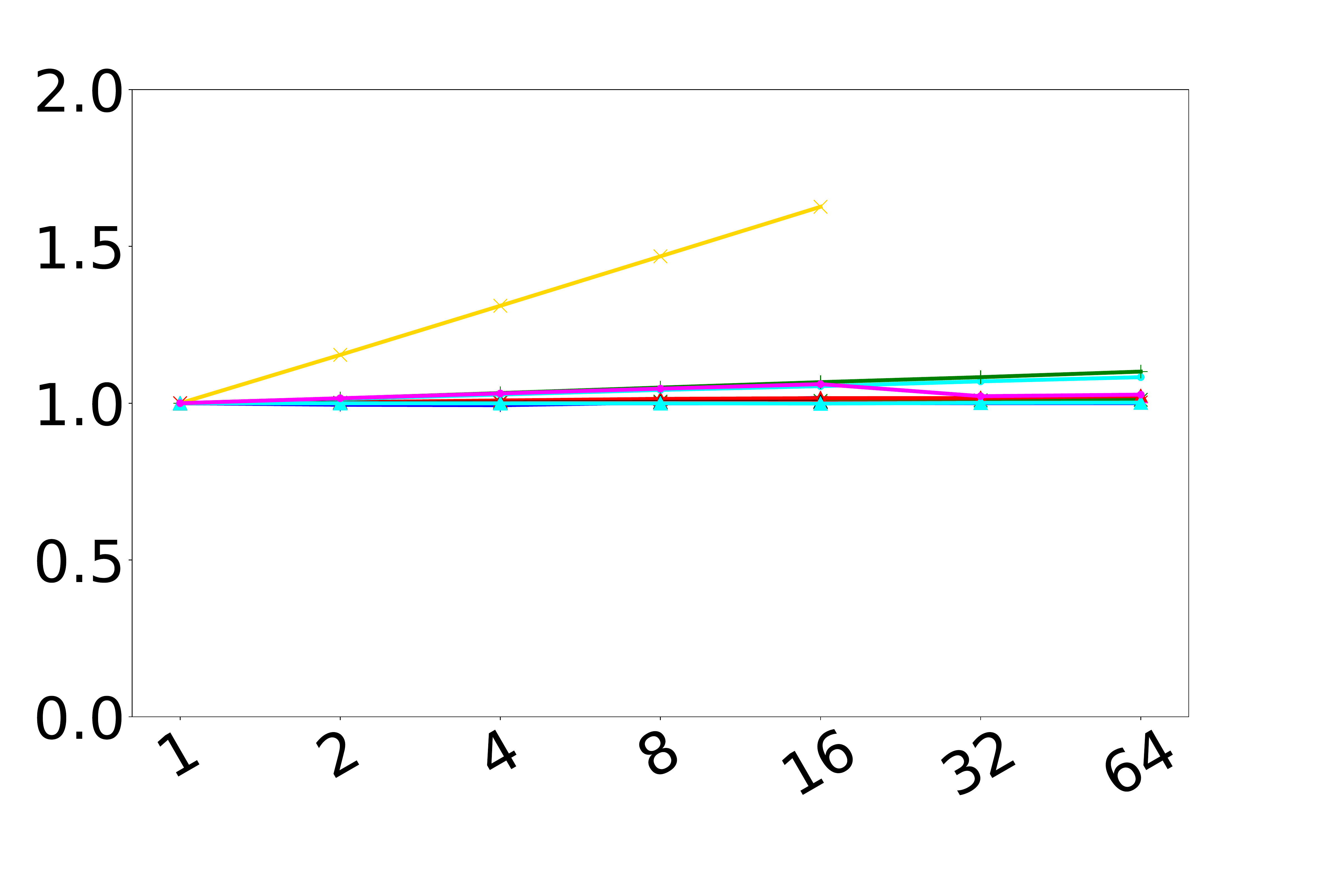}
		\caption{SM per cluster}
		\label{fig:dse_sm_clus-xavier}
	\end{subfigure}
	\begin{subfigure}[b]{0.32\textwidth}
		\includegraphics[width=\textwidth]{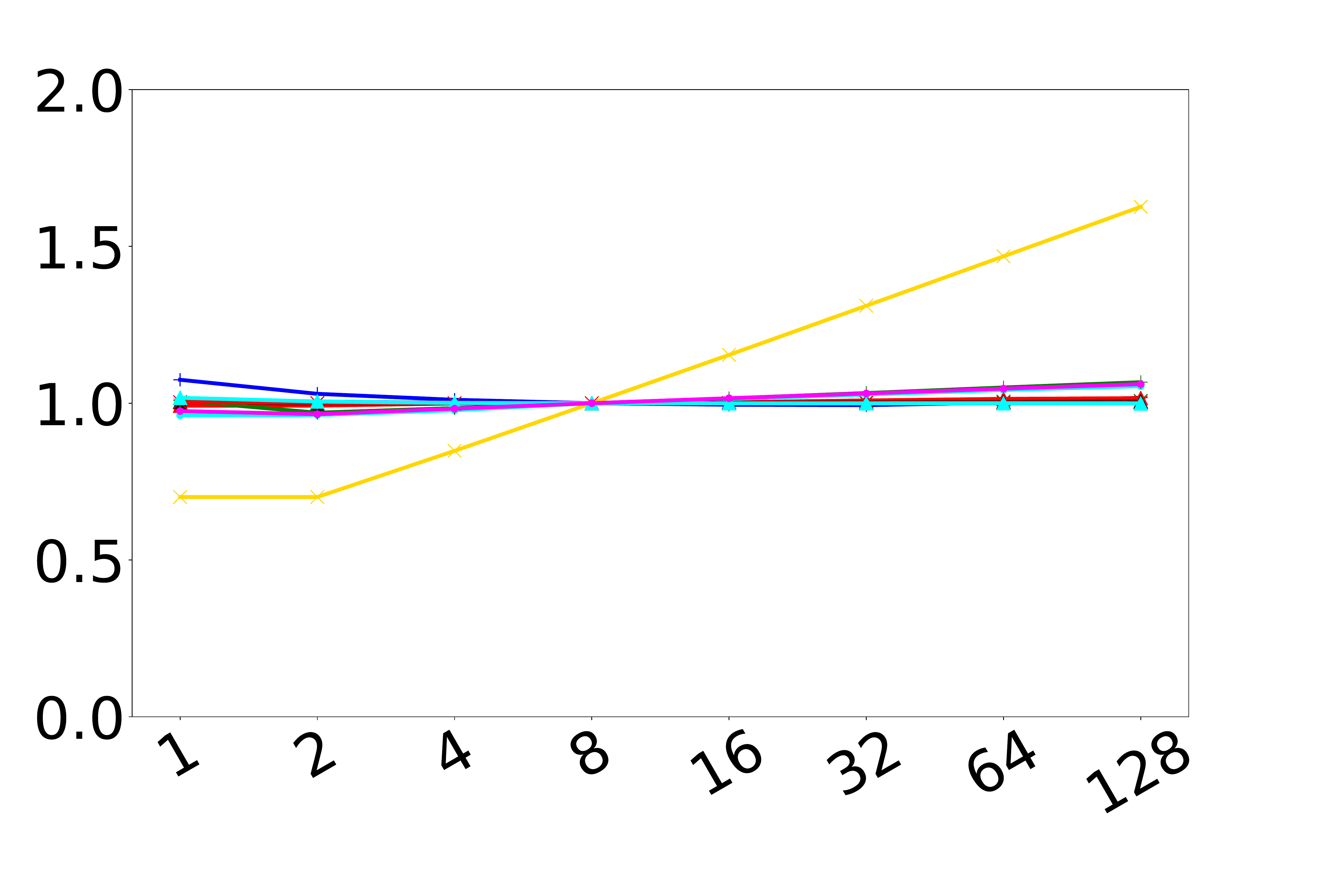}
		\caption{Number of SMs}
		\label{fig:dse_cluster-xavier}
	\end{subfigure}
	\begin{subfigure}[b]{0.32\textwidth}
		\vspace{-10em}
		\includegraphics[width=\textwidth]{img/plots/single/dse_legend.pdf}
		\caption{Legend}
		\label{fig:dse_legend-xavier}
	\end{subfigure}
	
	\caption{Design Space Exploration results for one parameter changes in the AGX Xavier (all sizes are in KB).}
	\label{fig:dse-xavier}
\end{figure*}

\begin{figure*}[t!]
	\centering
	
	\begin{subfigure}[b]{0.32\textwidth}
		\includegraphics[width=\textwidth]{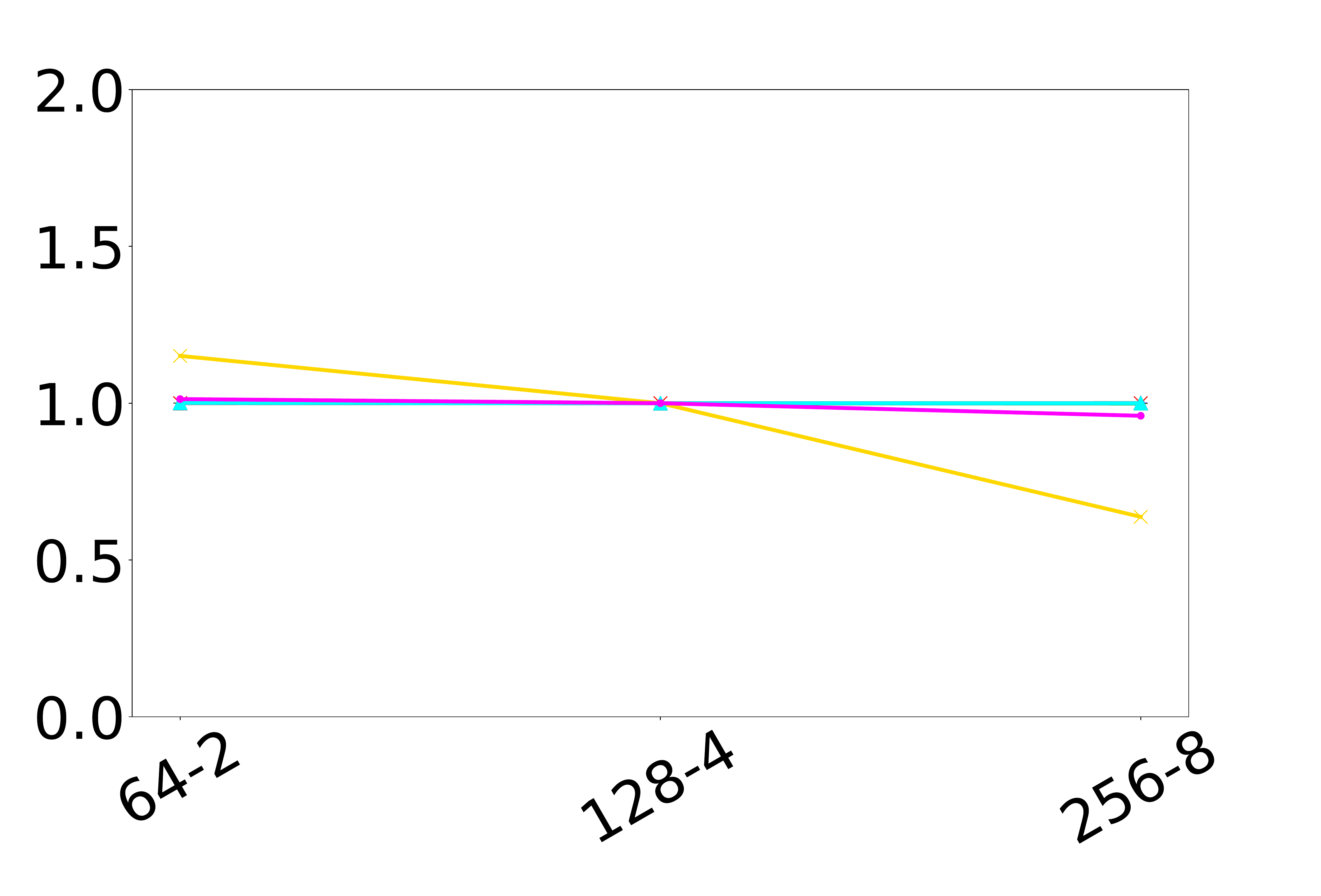}
		\caption{L1 size and associativity}
		\label{fig:dsem_l1_sz_as-xavier}
	\end{subfigure}
	\begin{subfigure}[b]{0.32\textwidth}
		\includegraphics[width=\textwidth]{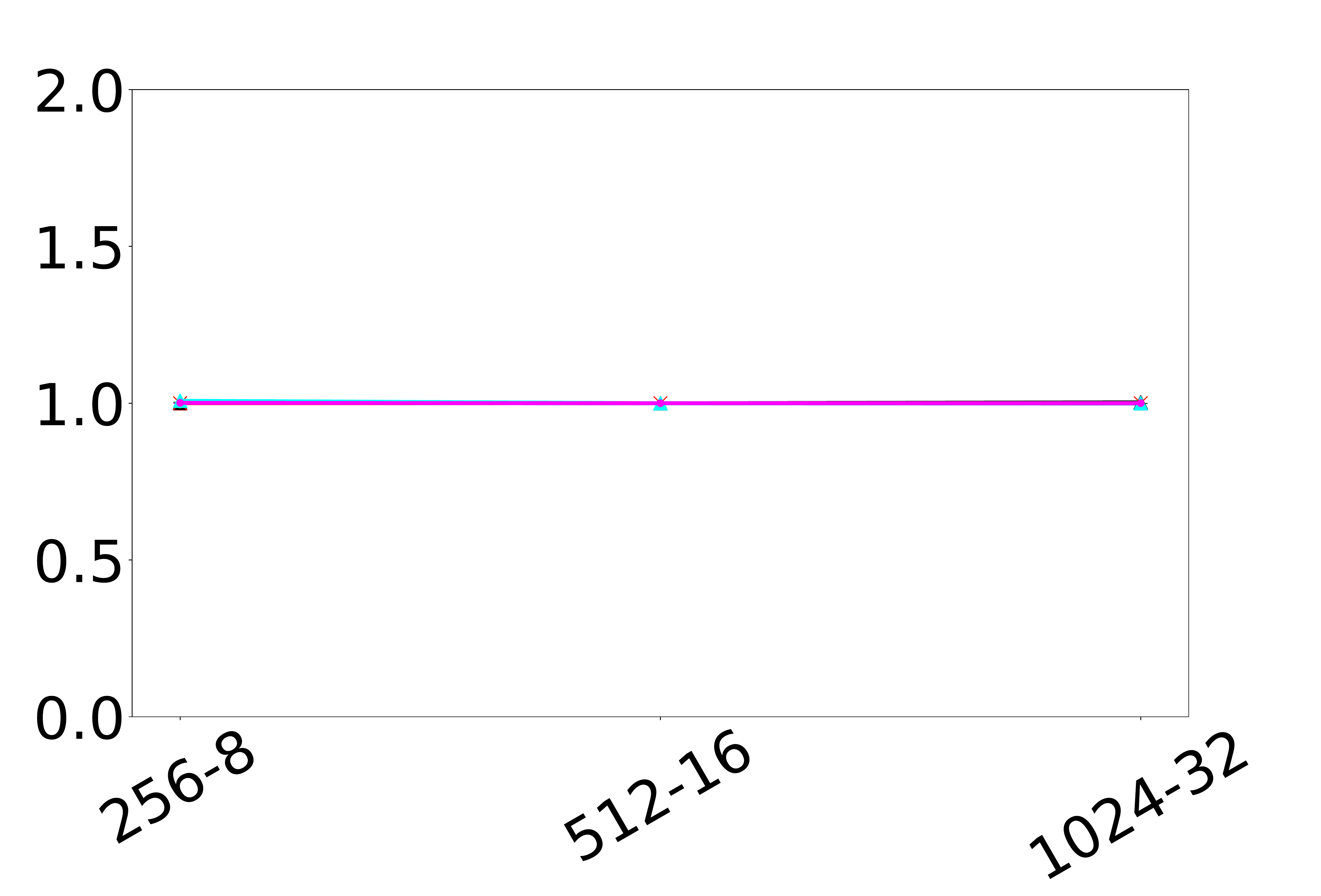}
		\caption{L2 size and associativity}
		\label{fig:dsem_l2_sz_as-xavier}
	\end{subfigure}
	\begin{subfigure}[b]{0.32\textwidth}
		\includegraphics[width=\textwidth]{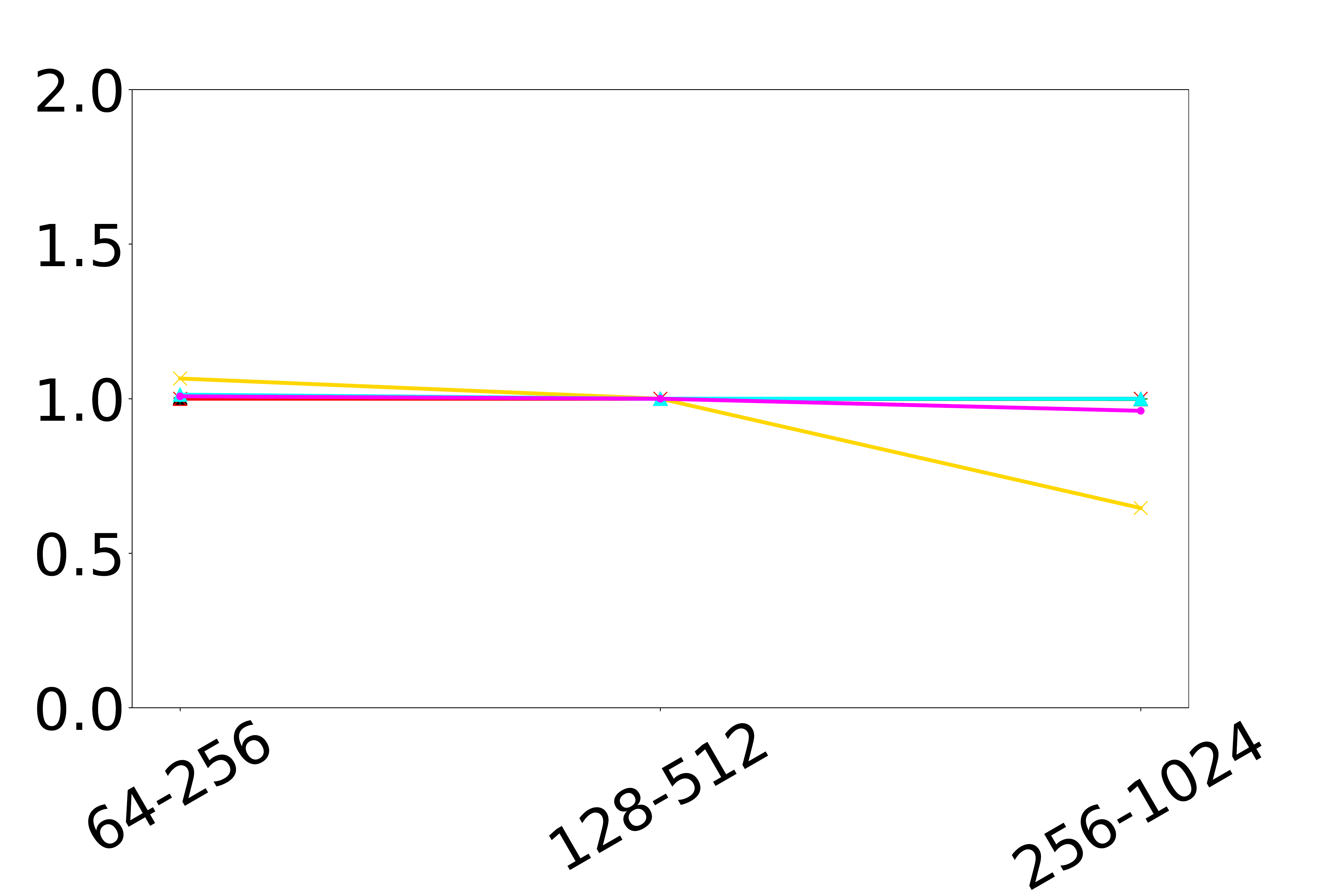}
		\caption{L1 and L2 size}
		\label{fig:dsem_l1_l2_sz-xavier}
	\end{subfigure}
	
	\begin{subfigure}[b]{0.32\textwidth}
		\includegraphics[width=\textwidth]{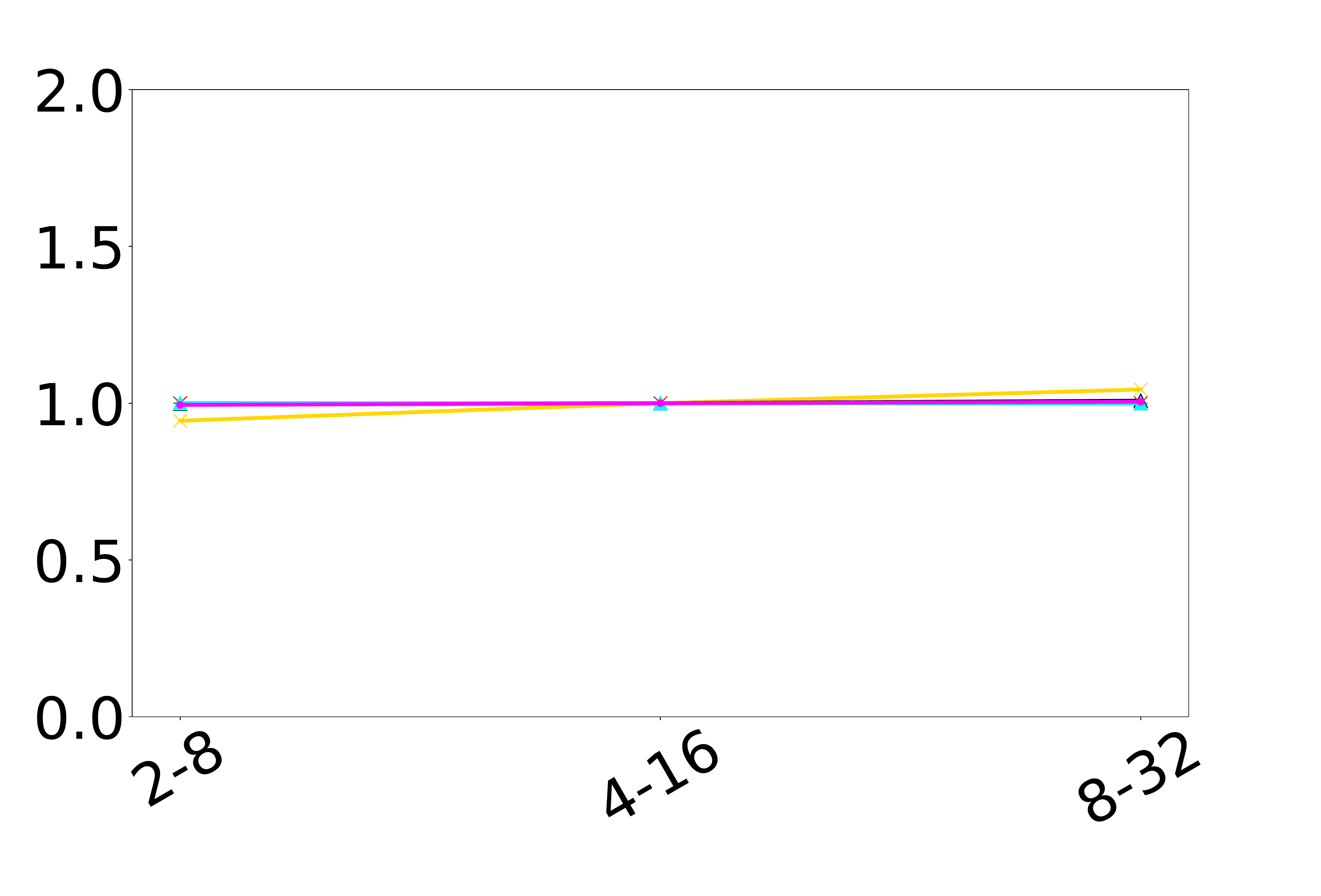}
		\caption{L1 and L2 associativity}
		\label{fig:dsem_l1_l2_as-xavier}
	\end{subfigure}
	\begin{subfigure}[b]{0.32\textwidth}
		\includegraphics[width=\textwidth]{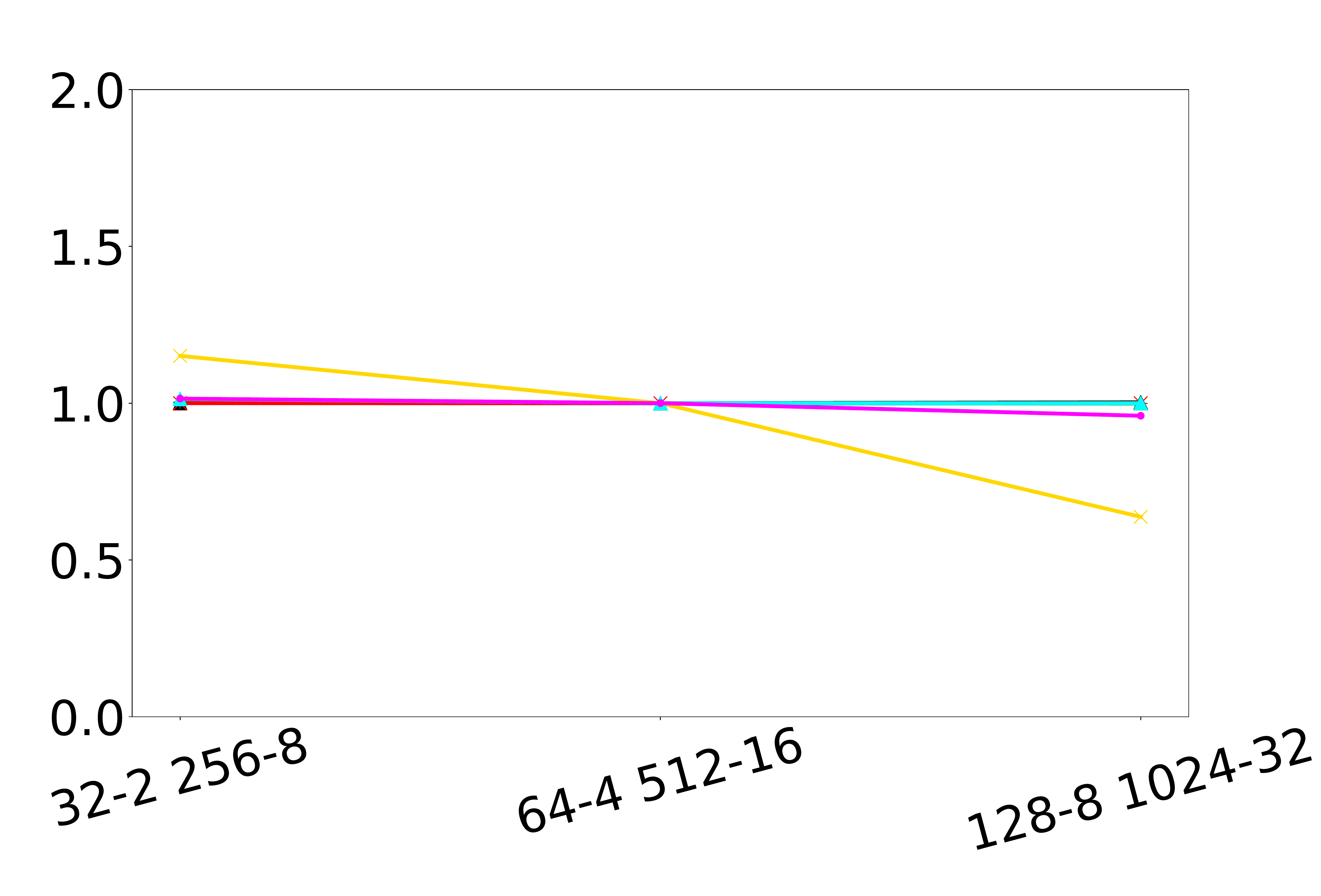}
		\caption{L1 and L2 size and associativity}
		\label{fig:dsem_l1_l2_sz_as-xavier}
	\end{subfigure}
	\begin{subfigure}[b]{0.32\textwidth}
		\includegraphics[width=\textwidth]{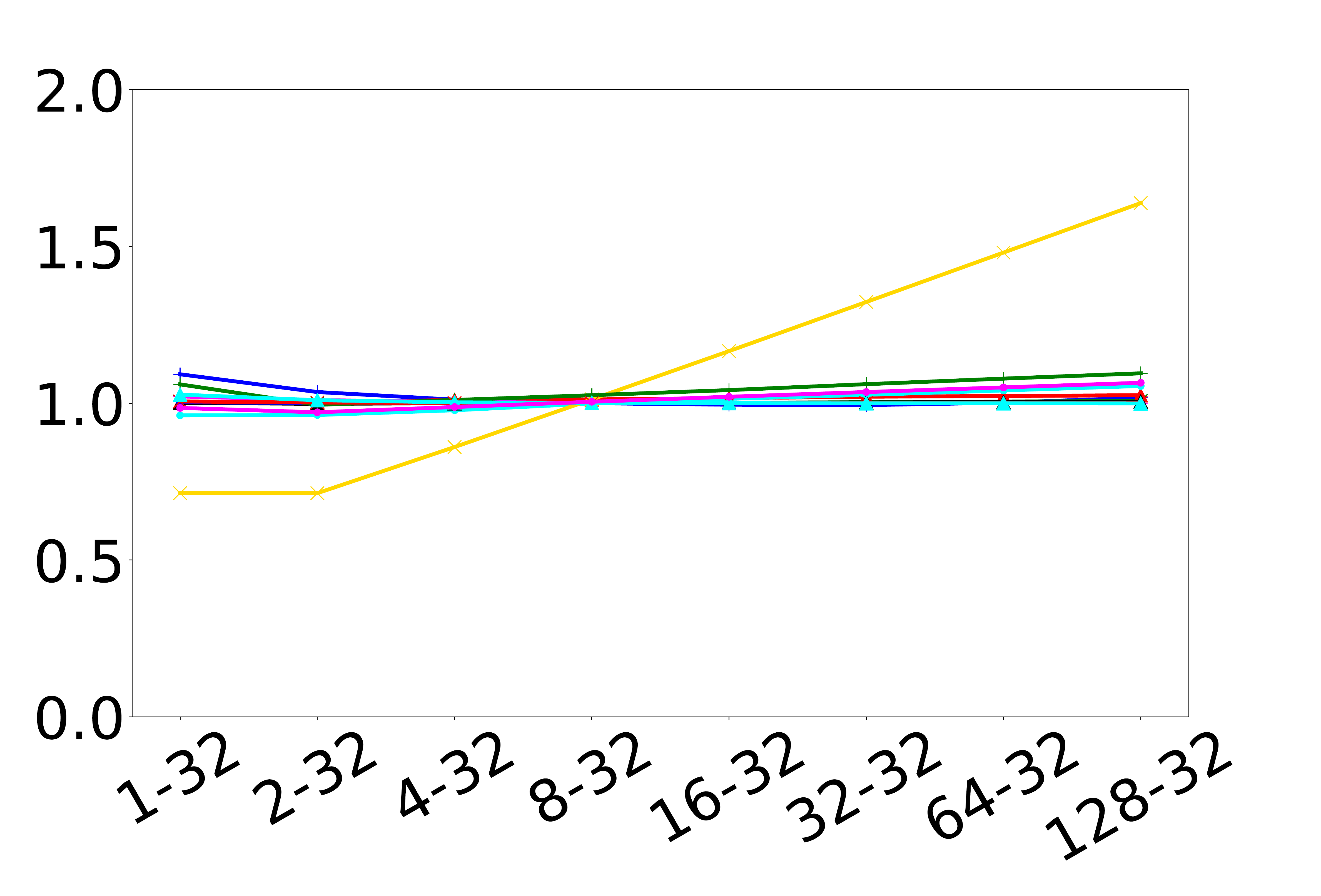}
		\caption{SM with 32 CUDA cores}
		\label{fig:dsem_sm_32-xavier}
	\end{subfigure}
	
	\caption{Design Space Exploration results for several parameter changes in the AGX Xavier (all sizes in KB). The legend is the same used in Figure~\ref{fig:dse-xavier}.}
	\label{fig:dsem-xavier}
\end{figure*}

\subsection{AGX Xavier: Changing a single parameter}

In Figure~\ref{fig:dse-xavier}, we see the results for the design space exploration when changing only one parameter.
Y-axis shows the slowdown with respect to the baseline configuration.
The first 4 plots show the results for the variation in L1 size, L1 associativity, L2 size, and L2 associativity. Increasing the L1 size provides a small performance improvement for some of the benchmarks, while reducing the size degrades performance. As shown, one of the benchmarks (\textit{myocyte}) is the most sensitive one to the L1 size. When reducing the L2 cache size, we observe performance degradation, however, further increasing the L2 cache size does not provide further performance improvement, similar to the case of TX2.

Regarding the associativity, neither changes in L1 nor in L2 result in significant changes when increasing or decreasing it moderately (1 to 64 ways). These parameters are again tested in the next section since changing both size and associativity at the same time may have different effects that are not seen when changing them one at a time.

Figure~\ref{fig:dse-xavier} e) shows that doubling the number of CUDA cores does not increase performance, however, unlike the TX2, reducing it does not cause significant performance degradation. In Figure~\ref{fig:dse-xavier} f), we show the effects of changing the size of the register file. As shown, some benchmarks are sensitive when reducing the register file size, however, the rest show very low sensitivity. Figure~\ref{fig:dse-xavier} g) changes the number of warp schedulers in each SM. Again, unlike the TX2, benchmarks show very low sensitivity to the change of number of warp schedulers. As it shows, increasing the number of warp schedulers from 1 to 16 has almost no impact. 

In Figures~\ref{fig:dse-xavier} h),~\ref{fig:dse-xavier} i), and~\ref{fig:dse-xavier} j), we show the results for the number of SMB per SM, the number of SM per cluster, and the number of clusters with just 1 SM, respectively. As explained earlier, the simulator does not support more than 4 SMB per SM (the standard in Pascal). Therefore, we just focus on reducing it to 1 and 2. We observe that this reduction would incur in very small performance loss unlike the case of TX2. The last two (SM per cluster and number of clusters) show a similar trend, since, in the end, both of them are increasing the total amount of SMs. We observe that increasing the total number of SMs has positive effects in performance. Some benchmarks, such as \textit{myocyte}, can benefit the most from increasing the number of SMs and also the number of SMs per cluster. As discussed earlier, the increase in SMs has to be tailored to the application implementation and behavior, so some applications may not be able to use all these SMs while others could if they were implemented with more granularity. These two components are the ones that vary most depending on the benchmark, with several improving performance and others decreasing it.

\subsection{AGX Xavier: Changing two or more parameters}

Similar to the analysis that we have shown for the TX2, some parameters that changed on their own have a specific impact on performance, can behave differently if changed together with other parameters. This is because how one parameter performs depends also on how other parameters perform. This is obvious in caches. Changing the associativity of a small cache may have bigger impact on performance than changing the associativity of a big cache, since the number of sets may be large enough to mitigate cache conflicts.

The first two parameters that we have changed together are cache size and associativity, both for L1 (Figure~\ref{fig:dsem_l1_sz_as-xavier}) and L2 (Figure~\ref{fig:dsem_l2_sz_as-xavier}). For both, we have changed the associativity (first element in the x-axis) and the size (second element) together. Although for the L2 we do not see significant performance changes, for the L1 we see that, by reducing the size and associativity to small enough numbers, we have  performance degradation for one of the sensitive benchmarks. 

In the next experiments, we change the size of both caches at the same time (Figure~\ref{fig:dsem_l1_l2_sz-xavier}), and we see that increasing both cache sizes together can improve performance in some benchmarks. {\color{black}However, performance gains due to increasing L1 and L2 sizes simultaneously are similar to those of increasing L1 size only, which indicates that there is no particular combined effect when increasing L1 and L2 sizes simultaneously.} We also changed both L1 and L2 associativities at the same time (Figure~\ref{fig:dsem_l1_l2_as-xavier}) and, as in previous experiments, we see that it has no noticeable effect. In the last experiment with multiple cache components, we change both L1 and L2 sizes and associativities. The results are similar to b) and c), with small performance penalties for one of the benchmarks in the small setup and also significant performance improvement in the big one.

Finally, in Figure~\ref{fig:dsem_sm_32-xavier}, we change both the number of SM as well as the number of CUDA cores per SM, changed from 128 to just 32. Comparing with the previous experiment, Figure~\ref{fig:dse_cluster-xavier}, we see that the trends are similar, with small variations depending on the benchmark. The \textit{myocyte} benchmark, however, is highly impacted as expected{\color{black}, but performance degradation does not differ significantly from that when using 128 instead of 32 CUDA cores per SM}.

}

\subsection{Changing the software}

The changes in performance that we observe when modifying different hardware parameters not only depend on the type of application we are running (compute-intensive, memory-intensive, etc.) but also on the specific CUDA/OpenCL implementation of each application. Usually, when parallelizing an application for GPUs, the computation is divided into grids and thread blocks (using NVIDIA's terminology). A specific grid and thread block division of a program could be optimal for a configuration (number of SM, sizes of caches etc) while being suboptimal for others.

Because of this, we have modified the implementation of the programs to change its granularity, to see the impact that these can have with different hardware setups.
As a proof of concept, we changed the grid and thread block distribution of 3 benchmarks: cell, needle and nn. Since sometimes this distribution is tightly related with the specific implementation, changing the code to accommodate the new distribution can be challenging.

In order to test whether this new software distribution with more potential parallelism is utilizing the added hardware, we test two configurations. Both these configurations have bigger L1 and L2 caches (364KB and 2048KB respectively), and they differ in the distribution of SM and CUDA cores. The first one has 16 SMs and 128 CUDA cores per SM and the second one 16 SMs and 32 CUDA cores per SM. The reference one has 2 SMs with 128 CUDA cores per SM.

\subsection{Parameter classification}

Depending on the results obtained in the design space exploration, we classify the parameters into two categories:

\begin{enumerate}

\item Parameters that (based on Rodinia) are not worth to increase beyond a given point since they produce no gains.

\item Parameters that require SW to be modified (e.g. block/grid) to make it worth to change them.

\end{enumerate}

In the following table (Table~\ref{tb:class}) for the TX2, we show a classification of the parameters into the two different types, and the last column shows the limit where an increase of the parameter does not provide significant performance benefits.

\begin{table}[]
	\centering
	\begin{tabular}{|l|cr|}
		\hline
		\multicolumn{1}{|c|}{\textbf{Parameter}} & \multicolumn{1}{c}{\textbf{Category}} & \multicolumn{1}{c|}{\textbf{Limit}} \\
		\hline
		L1 size & 1 & 48KB \\
		L1 associativity & 1 & 4 \\
		L2 size & 1 & 256KB \\
		L2 associativity & 1 & 2 \\
		Number of CUDA cores & 1 & 128 \\
		Register file size & 1 & 64KB \\
		Shared memory size & 1 & 16KB\\
		Number of warp schedulers & 1 & 2 \\
		Number of SMB per SM & 2 & -\\
		SM per cluster & 2 & -\\
		Number of SMs & 2 & -\\
		\hline
	\end{tabular}

	\caption{TX2 GPU parameters classified depending on the potential improvement on hardware or software.}
	\label{tb:class}
\end{table}

%% file: 5.0.Improvements.tex
\section{Improved setups}
\label{sec:improved}

Based on the knowledge obtained from the design space exploration, we want to propose modified hardware designs with some improvement, either in terms of performance or in terms of die area. The two setups that we propose are:

\begin{itemize}
	\item The same performance with less cost.
	\item More performance with the same cost.
\end{itemize}

These setups would be optimized for the Rodinia benchmarks used in the design space exploration and will be based on the TX2 and AGX Xavier basic design.

\subsection{Proposed setups}

The two proposed setups and its objectives are:

\begin{itemize}
	\item \textit{Decrease the hardware keeping the same performance.}

The first improved setup proposed aims to reduce the amount of die space used while roughly keeping the same performance (within 5\% of performance degradation).

For the case of TX2, looking at the results of the design space exploration, the features that are susceptible to being reduced without too much performance degradation are: register size, warp schedulers per SM and shared memory. In all of these features, the number of units or size can be divided by half without having a significant impact in performance.

Thus, our proposal is to use the same configuration but reducing the register size from 64KB to 32KB ($\frac{1}{2}$), the warp schedulers per SM from 4 to 2 ($\frac{1}{2}$) and the shared memory from 64KB to 16KB ($\frac{1}{4}$).

{\color{black}In the case of the AGX Xavier, following analogous reasoning, we propose to halve the number of warp schedulers (from 4 to 2), as well as the size of the register file (from 64K to 32K).}

	\item \textit{Increase performance using the same hardware.}

Again, in the case of TX2, in order to increase the performance using the same amount of die space, we need to increase some resources, which will increase performance and area and decrease others, which will decrease performance but decrease area. The trade-off between the resources increased and decreased needs to be positively balanced to improve the performance per die area.

Based on the design space exploration, our proposal for this setup is to increase the number of SMs and L1 cache sizes, while decreasing the size of the L2. The difference in die space between the new SMs and larger L1 should be similar to the decrease in L2 size. Furthermore, we will provide two variations of this setup. In one we will double the size of the L1 and the number of SMs, while reducing the L2 size by half, as Table~\ref{tb:improved-tx2} shows. Furthermore, in a more area limited setup, we will also double the L1 and number of SMs, while reducing the L2 to one fourth of its size.

{\color{black}Regarding the AGX Xavier, following analogous reasoning, we propose increasing the number of SMs (from 8 to 16), at the expense of decreasing L2 cache size down to 256KB or 128KB.}

\end{itemize}

\begin{table}[]
	\centering
		\begin{tabular}{|l|rrrr|}
			\hline
			\multicolumn{1}{|c|}{\textbf{Parameter}} & \multicolumn{1}{c}{\textbf{Base}} & \multicolumn{1}{c}{\textbf{Reduced}} & \multicolumn{1}{c}{\textbf{Increased}} & \multicolumn{1}{c|}{\textbf{Increased}} \\
			\multicolumn{1}{|c|}{\textbf{changed}} & \multicolumn{1}{c}{\textbf{setup}} & \multicolumn{1}{c}{\textbf{die space}} & \multicolumn{1}{c}{\textbf{perf. a)}} & \multicolumn{1}{c|}{\textbf{perf. b)}} \\
			\hline
			Number of SM & 2 & 2 & 4 & 4 \\
			Warp sched & 4 & 2 & 4 & 4 \\
			Register file & 64KB & 32KB & 64KB & 64KB \\
			Shared mem & 64KB & 16KB & 64KB & 64KB \\
			L1 size & 48KB & 48KB & 96KB & 96KB \\
			L2 size & 512KB & 512KB & 256KB & 128KB \\
			\hline
		\end{tabular}
	\caption{Changes made in the improved setups for the TX2.}
	\label{tb:improved-tx2}
\end{table}

\begin{figure}[h]
	\centering
	\includegraphics[width=\columnwidth]{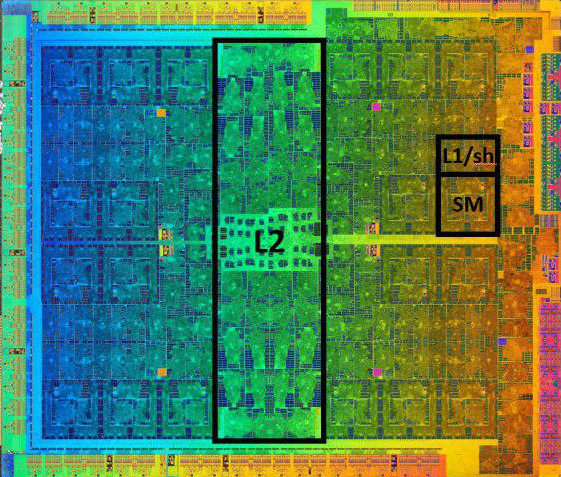}
	\caption{NVIDIA GTX 1080 die. Same architecture (Pascal) and manufacturing process (16nm) as the TX2.}
	\label{fig:1080-die}
\end{figure}

\begin{table}[]
	\centering
	\begin{tabular}{|l|rrrr|}
		\hline
		\multicolumn{1}{|c|}{\textbf{Parameter}} & \multicolumn{1}{c}{\textbf{Base}} & \multicolumn{1}{c}{\textbf{Reduced}} & \multicolumn{1}{c}{\textbf{Increased}} & \multicolumn{1}{c|}{\textbf{Increased}} \\
		\multicolumn{1}{|c|}{\textbf{changed}} & \multicolumn{1}{c}{\textbf{setup}} & \multicolumn{1}{c}{\textbf{die space}} & \multicolumn{1}{c}{\textbf{perf. a)}} & \multicolumn{1}{c|}{\textbf{perf. b)}} \\
		\hline
		Number of SM & 8 & 8 & 16 & 16 \\
		Warp sched & 4 & 2 & 4 & 4 \\
		Register file & 64KB & 32KB & 64KB & 64KB \\
		L1 size & 128KB & 128KB & 256KB & 256KB \\
		L2 size & 512KB & 512KB & 256KB & 128KB \\
		\hline
	\end{tabular}
	\caption{Changes made in the improved setups for the AGX Xavier.}
	\label{tb:improved-xavier}
\end{table}

Justifying that the hardware cost of our proposed setups is challenging without having information about the actual space occupied by each resource in the real hardware implementation. Since there is no available information about the TX2 die, instead we use, as a reference, the GTX 1080's die information. The GTX 1080 is a discrete graphics card developed by NVIDIA with the same architectural generation (Pascal) and manufacturing process (16nm).

In the GTX 1080 die, we see the area dedicated to the SMs to the per SM caches (including L1 and shared memory), to the shared L2 and the rest to I/O and other features (debugging, performance counters etc). Although these are estimates, they give us a good idea of the overall die space used to each processor part.

\begin{figure}[t]
	\centering
	\includegraphics[width=\columnwidth]{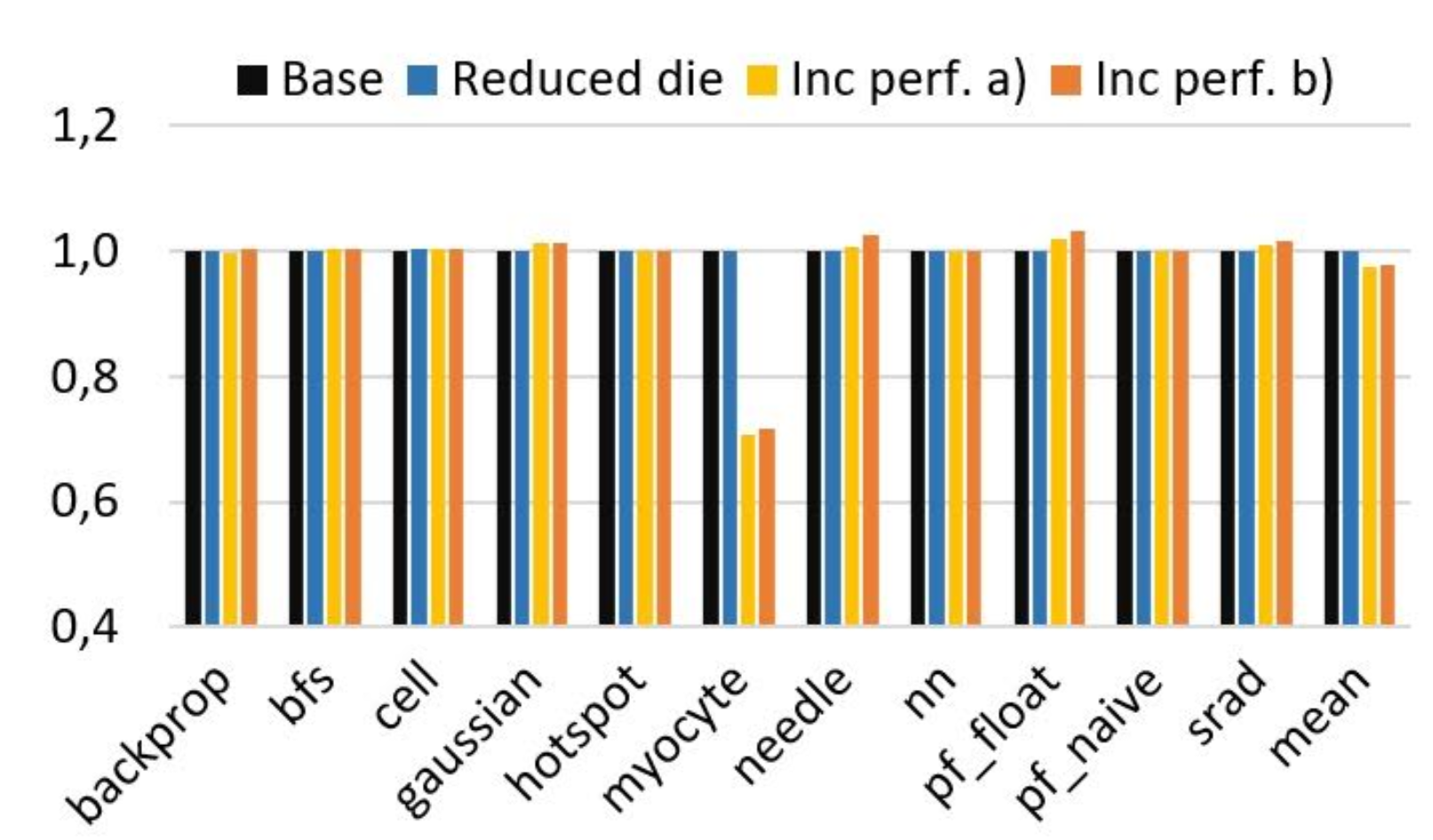}
	\caption{Execution time of the proposed setups normalized to the baseline for the AGX Xavier.}
	\label{fig:improved-xavier}
\end{figure}

{\color{black}
We have proposed similar changes to be studied on the parameters of the AGX Xavier, as discussed before. Table~\ref{tb:improved-xavier} shows the changes made in different parameters in the improved setups for the AGX Xavier.

\subsection{Evaluation}

Regarding the TX2, in the first setup, the improvement is in the die space used. Specifically, we use half the register size, half the schedulers and a quarter of the shared memory. Of the three, the shared memory is the one that has a bigger impact in die area reduced. {\color{black}An advantage of reducing these components is that they are private to each SM, so we could increase the number of SMs in combination with those SM-local component reductions without increasing overall area. In contrast, by decreasing shared resources such as the L2 cache, the effect is not multiplicative as for intra-SM components.}

From the 11 Rodinia benchmarks that we analyzed, 8 of them have the same performance (within less than 1\% of variation). Based on our previous analysis, from the 3 benchmarks that show a significant increase in execution time, one (hotspot) is mainly due to the decrease in register file size and the other 2 (needle and srad) are due to the decrease in shared memory size. On average, less than 5\% of performance degradation is shown when making these changes as Figure~\ref{fig:improved-tx2} shows.

In the second setup, we want to improve performance while increasing some resources and decreasing others to maintain die area. Since knowing the exact area gained or lost with each change is challenging, we propose two setups. In both, we double the number of SMs and the L1 size, but in one, we reduce the L2 size by half and in the other, by one quarter.

For both variations, the results change significantly depending on the benchmark. Of the 11 benchmarks, in 4 of them, we observe a performance improvement, in 4 of them we observe a slowdown, and the rest stay roughly the same. Overall, the setup that halves the L2 shows a 15\% reduction in execution time while the setup that divides it by 4 shows a 9\%.

Between the two setups, some benchmarks (bfs, cell, gaussian, hotspot, particlefiler\_naive) show no difference at all. Of the ones that show difference, all perform better in the 256KB L2 than in the 128KB.

\begin{figure}[t]
	\centering
	\includegraphics[width=\columnwidth]{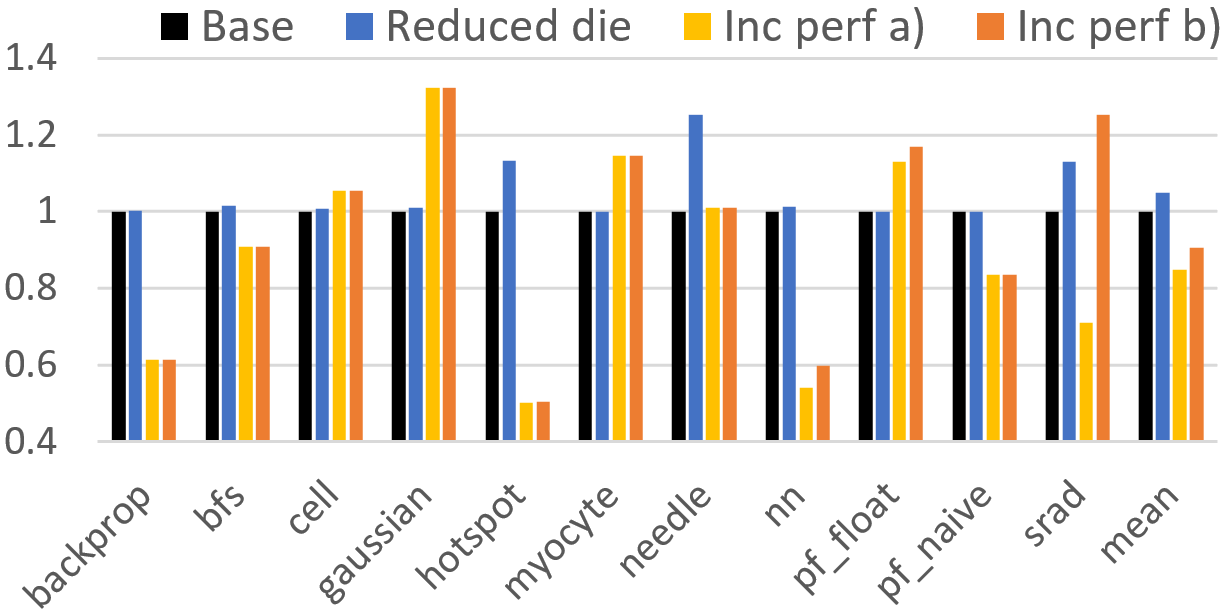}
	\caption{Execution time of the proposed setups normalized to the baseline for the TX2.}
	\label{fig:improved-tx2}
\end{figure}

Regarding the Xavier, as Figure~\ref{fig:improved-xavier} shows, the benchmarks are less affected by modifying the parameters. Our conclusion is that most of the benchmarks are less sensitive to some key parameters. However, few benchmarks, such as \textit{myocyte}, highly benefit from the improved setups as the Figure shows.
}

%% file: 6.0.Related.tex
\section{Related work}
\label{sec:related}

GPU performance analysis for general purpose applications has been a research topic for many years. Two main research lines exist on this topic. The first one focuses on the analysis of Commercial Off-The-Shelf (COTS) GPUs by means of the execution of different types of parallel applications~\cite{COTSHPC1,COTSHPC2,COTSHPC3}. These studies have allowed determining what the most convenient way is to deploy and run software on those GPUs and, at most, it has been guessed how hardware should be modified to improve performance. However, since COTS GPUs cannot be modified, hypotheses raised cannot be verified.
The second research line on this topic considers the use of GPU performance simulators in order to determine how to tune high-performance GPUs for general purpose high-performance applications~\cite{simHPC1,simHPC2}. However, those GPUs are significantly different from automotive ones since they are not subject to strict power constraints such as those in the automotive domain, which are intended to operate under much lower power envelops. Thus, conclusions cannot be extrapolated across both market segments.

Several works have analyzed the performance of COTS automotive GPUs to optimize the behavior of applications running atop~\cite{COTSauto1}. While conclusions reached by those works are highly valuable for an efficient use of hardware, they do not provide any insight on how to optimize hardware design.
Finally, some works target task scheduling on GPUs for an efficient use of hardware resources, minimizing their timespan while respecting their deadlines~\cite{GPUsched1,GPUsched2,GPUsched3}.

{\color{black}
In~\cite{mazzocchetti2019performance}, we performed an extensive analysis and design space exploration of the key GPU parameters for the NVIDIA TX2 GPU. In this paper, we complement such work by performing a similar study for a more recent and widely-used automotive GPU, the NVIDIA AGX Xavier. In addition, we discuss the differences among these two architectures and explain why in some case we observe different behavior while modifying similar parameter(s).
	
}

%% file: 7.0.Conclusions.tex
\section{Conclusions}
\label{sec:conclusions}

Performance requirements of future automotive systems have increased significantly with the promise of Autonomous Driving. GPUs are commonly used to reach the required performance levels. In this work, we focus on two of the latest automotive SoCs NVIDIA Jetson TX2 and NVIDIA AGX Xavier which are widely-used and well-known platforms.

First, we modeled both the TX2 and the AGX Xavier in our full-system CPU-GPU simulator. Then, using Rodinia benchmark suite that focuses on several heterogeneous computing (CPU+GPU) applications, we analyzed how different parameters of the GPU in the TX2 and AGX Xavier are affecting performance. This is done with isolated changes (just changing a single parameter at the same time) as well as with combined changes (changing several parameters at the same time).

Building on the conclusions of this experimentation, we propose three different improved setups for each of the GPUs: one that reduces die space significantly with less than 5\% performance impact, and two that maintain approximately the same die space but reduce execution time between 5\% and 15\%.

Some hardware improvements could only be fully exploited by modifying the software (mainly the block and thread distribution). We leave this tuning of software as future work to continue improving GPU-based systems to meet the demands of the automotive industry.